\documentclass[11pt]{article}
\usepackage{epsf,epsfig,float,amssymb,latexsym}
\usepackage{graphics,psfrag}
\newcommand{\xxpc}{\%}

\newcommand{\vk}{{\mathbf{k}}}

\newcommand{\Opd}{{\mathcal{O}}(p^2)}

\newcommand{\Opc}{{\mathcal{O}}(p^4)}

\newcommand{\Ops}{{\mathcal{O}}(p^6)}

\newcommand{\be}{\begin{equation}}
\newcommand{\ee}{\end{equation}}
\newcommand{\ba}{\begin{eqnarray}}
\newcommand{\ea}{\end{eqnarray}}

\newcommand{\nn}{\nonumber}

\newcommand{\krig}[1]{\stackrel{\circ}{#1}}
\newcommand{\vs}{\vspace{-0.0cm}}
\newcommand{\la}{\langle}
\newcommand{\ra}{\rangle}
\newcommand{\mmp}{M_P}
\newcommand{\mmq}{M_Q}
\newcommand{\epe}{\varepsilon}
\newcommand{\mkp}{\krig{M}_\pi}
\newcommand{\mkk}{\krig{M}_K}
\newcommand{\mke}{\krig{M}_\eta}
\newcommand{\mkq}{\krig{M}_Q}

\begin{document}

\thispagestyle{empty}

\vspace{2cm}

\begin{center}
{\Large{\bf Non-Perturbative Study of the Light Pseudoscalar Masses in Chiral Dynamics}}
\end{center}
\vspace{.5cm}

\begin{center}
{\Large Jos\'e A. Oller\footnote{email: oller@um.es} and Luis Roca\footnote{email: luisroca@um.es}}
\end{center}

\begin{center}
{\it  Departamento de F\'{\i}sica. Universidad de Murcia.\\ E-30071
Murcia.  Spain.}
\end{center}
\vspace{1cm}

\begin{abstract}

\noindent
We perform a non-perturbative chiral study of the masses of the lightest
pseudoscalar mesons. In the calculation of the self-energies 
 we employ the S-wave meson-meson amplitudes taken from
 Unitary Chiral Perturbation Theory (UCHPT) that include the lightest nonet
  of scalar resonances.  Values for the
bare masses of pions and kaons are obtained, as well as  an estimate
of the mass of the $\eta_8$. The former are found to dominate the physical pseudoscalar masses.
 We then match to the self-energies
from Chiral Perturbation Theory (CHPT) to  $\Opc$, and a robust relation between 
several $\Opc$ CHPT counterterms is obtained. We also 
resum higher orders from our calculated  self-energies. By taking into account 
values determined from previous chiral phenomenological studies of 
$m_s/\hat{m}$ and $3L_7+L^r_8$, we determine a tighter region of favoured 
 values for the $\Opc$ CHPT counterterms $2L^r_6-L^r_4$ and 
 $2L^r_8-L^r_5$. This determination perfectly overlaps with the recent determinations
  to $\Ops$ in CHPT.  We warn about a likely
  reduction in the value of $m_s/\hat{m}$ by higher loop diagrams and that 
 this is not systematically accounted for by present lattice extrapolations. We also 
 provide a favoured interval of values for $m_s/\hat{m}$ and $3L_7+L^r_8$.
\end{abstract} \newpage

\section{Introduction}
\label{sec:intro}
\def\theequation{\arabic{section}.\arabic{equation}}
\setcounter{equation}{0}

Chiral Perturbation Theory (CHPT) is the effective quantum field
theory of strong interactions  at low energies
\cite{wein,gl1,glsu3}.  Three flavour CHPT, including strangeness,  has
already reached a very sophisticated stage  with present
calculations at the level of next-to-next-to-leading order (NNLO) or
$\Ops$  \cite{8bij,9bij,bijtalk}. In ref.\cite{bijtal,mass}
the masses and decay constants  of the lightest octet of
pseudoscalar mesons are worked out at to $\Ops$. In this reference it is shown 
 that the $\Ops$ two loop contributions  to the self-energies of the pseudoscalars
  are much larger than the full $\Opc$. 
In addition,  the $\Ops$ tree level contributions to the kaon and eta masses, with the 
$\Ops$ chiral counterterms estimated by resonance saturation,   are
 much larger in  modulus  than the $\Ops$ pure
loop ones and with opposite sign. As stated in ref.\cite{mass} the
$\Ops$ counterterms ``seem severely  overestimated" due to the complicated nature
of the scalar sector. Hence, this
reference  rises the interesting question of how one can improve
the calculation of the  contributions from the lightest scalar resonances.
 It is
well known that the scalar sector is a source of large higher
order corrections to the CHPT series. As a relevant example, let us
quote  the  $I=0$ $\pi\pi$ scattering length, $a_0^0$, which receives
 large higher order corrections  (a $40$\% correction with
respect to the lowest order value). This should be compared 
with the {\it non-resonant} $I=2$ S-wave scattering
length, $a_0^2$, which has negligible higher order corrections
\cite{cola,french,ypela}.  Even larger higher order chiral
corrections due to the $I=0$ S-wave final state interactions happen
for the lowest order prediction of $\Gamma(\eta \to3\pi)$, more
than a factor of 2 \cite{leuty,truong}. Another good example
 is the process $ \gamma\gamma\to\pi^0\pi^0$ for which a 
  two-loop CHPT calculation is needed \cite{bellucci,sainionew} to improve the comparison with 
dispersive theory 
 calculations \cite{oroca,penmorgan,penanegra}.  In connection  with these
enhancements one has the $generation$ of the lightest
scalar resonances $\sigma$,  $f_0(980)$, $a_0(980)$ and $\kappa$
\cite{npa,nd,kpi,dobado} because of the self-interactions among the
pseudoscalars  in S-wave.

 The Feynman-Hellman theorem \cite{hellman} relates $\partial M^2/\partial m_q$, where 
 $M$ is a hadronic mass and $m_q$ the $q$ quark mass, with the 
$\bar{q}q$ scalar form factor of this hadron at zero four-momentum transfer, the so called
sigma-terms \cite{gassersigma}.  
 Of special relevance are those in connection with the lightest hadronic states, $\pi$, $K$,  
 $\eta$ for mesons and the nucleon for baryons. The large impact on the 
 evaluation of the pion-nucleon sigma-term of the $I=J=0$ $\pi \pi$ amplitude,
    due to the $t$-channel $\pi\pi$ exchange, is remarked in ref.\cite{sainio}. On the other hand, 
 it was stressed  in refs.\cite{gasmeis,ffgl,truong2,scalarpi}  that the scalar form factor 
 of the pion receives large unitarity corrections so that the one-loop CHPT approximation 
 becomes inaccurate at a surprisingly low energy. Here, the so called infrared singularities are
 enhanced in comparison
  e.g. with the vector form factor \cite{ffgl,nd}. 
 In ref.\cite{scalarpi} the scalar $K\bar{K}$ form factors were first evaluated 
at the one loop level in CHPT and also employed in UCHPT. The corrections to these form factors because
of the final state
interactions of the $I=J=0$ $\pi\pi$ and $K\bar{K}$ coupled channels were huge, see figs. 5 and 6 
of ref.\cite{scalarpi}. Since this happens to the scalar form factors one should expect a similar large
impact of the $I=J=0$ pseudoscalar-pseudoscalar amplitudes in evaluating the pseudoscalar masses, as
both quantities are related by the Feynman-Hellman theorem. We offer here 
an estimation of such effects, that includes the exchanges of the lightest scalar resonances
$\sigma$, $\kappa$, $f_0(980)$ and $a_0(980)$.

Recently, precise calculations of  the pseudoscalar masses in lattice QCD with three 
dynamical fermions are 
available \cite{milc,milc1,ultimo}. 
 In these evaluations lattice results are taken down to  the
 values of the lightest quark masses, $m_u$ and $m_d$, and extrapolated to 
the continuum  by
employing  perturbative SU(3) Staggered CHPT calculations \cite{9l,apco}. 
 Given the large size of the pure loop contributions at 
${\cal O}(p^6)$ estimated in  ref.\cite{mass}, and since they were  neglected
 in refs.\cite{milc1,ultimo}, one should consider the possibility whether 
 these large contributions are buried in the present values for quark masses and low energy 
parameters determined by this Collaboration.

The manuscript is organized as follows. Section \ref{sec:form} is
dedicated to the development  of the formalism for the calculation
of the self-energies. In section \ref{sec:dyn} the results without matching 
 with CHPT at
$\Opc$ are given, this is what  we call the full dynamical
self-energies. The $\Opc$ CHPT self-energies 
are introduced and supplemented with higher order corrections
 from our formalism in section \ref{sec:opc}. We also derive in this section a robust 
 relation between $L_{(4,6)}=2L^r_6-L^r_4$ and $L_{(5,8)}=2L^r_8-L^r_5$ and determine 
 our interval of values for these
 counterterms. It turns out that this region remarkably overlaps with the values determined in 
 refs.\cite{bijkpi,bijtal} from $\Ops$ fits to data. We discuss in section \ref{lattice} 
 about likely effects from higher loop  diagrams and determine the values 
of $m_s/\hat{m}$, $L_{(7,8)}=3L_7+L^r_8$,
 and self-energies from our favoured region of low energy chiral 
counterterms. Some conclusions are collected in 
 section \ref{sec:con}.

\section{Formalism}
\label{sec:form}
\def\theequation{\arabic{section}.\arabic{equation}}
\setcounter{equation}{0}

\begin{figure}[ht]
\psfrag{dela}{{${\delta^{(\hbox{o})}}$}}
\psfrag{ine}{$(1-\eta_{00}^2)/4$}
\psfrag{event}{Events}
\centerline{\epsfig{file=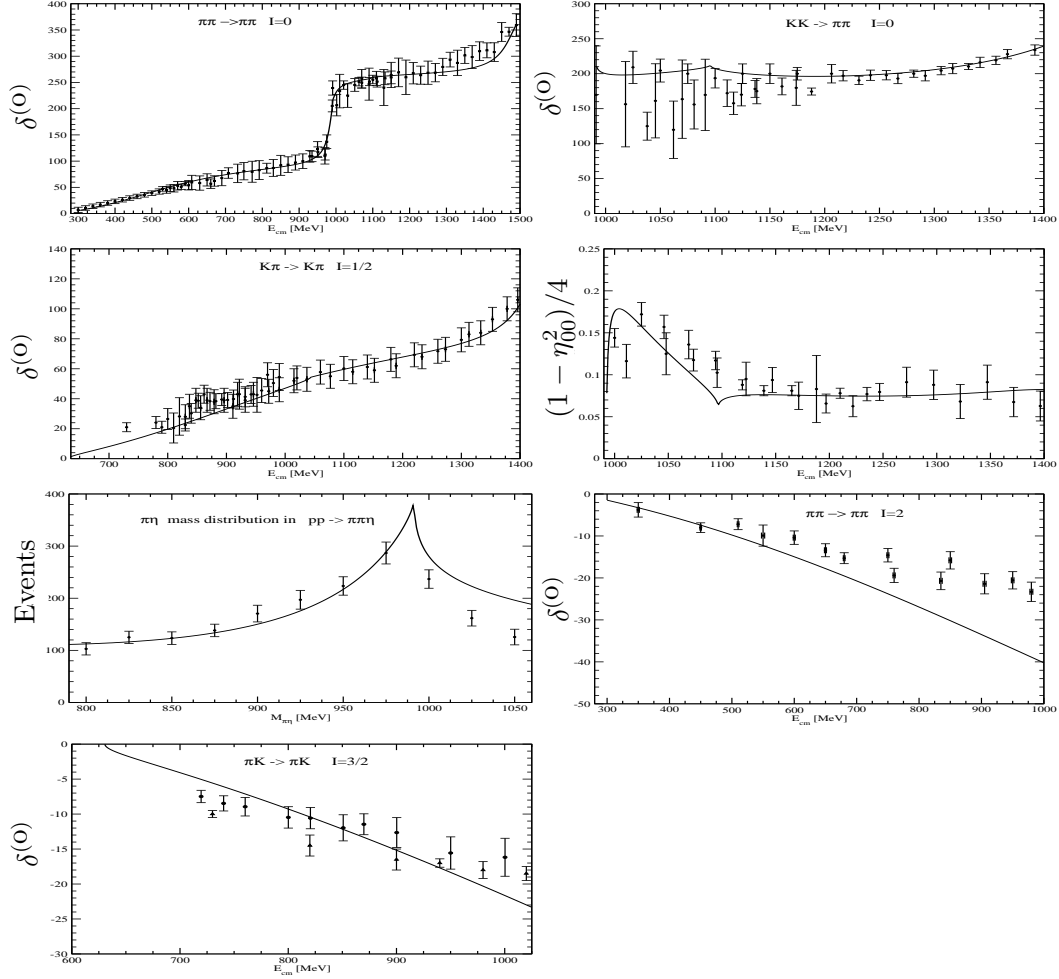,width=.8\textwidth,height=13cm,angle=0}}
\vspace{0.2cm}
\caption{\protect \small Different phase-shifts, inelasticities and mass
distributions obtained with the unitarized amplitudes. The experimental data 
are from
refs.~\cite{hyams,kaminski,ochs,frogratt,Cohen:1980cq,etkin,wetzel,mercer,
Estabrooks:1977xe,Arm,Linglin:1973ci,Martin:1979gm,PO38}
\label{fig:fitexp}}
\end{figure}

Our starting point is the S-wave meson-meson partial waves both for the resonant $I=0$, 1 and 1/2, 
 as well as for the much smaller and  non-resonant ones, $I=3/2$ and 2. 
  For the former set we take the amplitudes of ref.\cite{nd}.
The interaction kernel employed in this reference comprises the lowest order
CHPT amplitudes together with the s-channel exchange of  scalar resonances in a chiral symmetric 
invariant way  from ref.\cite{set}. These tree level resonances constitute  an octet with mass 
around 1.4~GeV and a singlet around 1~GeV. On the other hand, the coupled channels are $\pi\pi$, $K\overline{K}$ and $\eta\eta$ 
for $I=0$, $\pi\eta$ and $K\overline{K}$ for $I=1$ and $K\pi$ and $K\eta$ for $I=1/2$.
 In addition to the resonances explicitly
included at tree level, related to the physical ones around 1.4~GeV like e.g. the $K^*_0(1410)$ 
or the $a_0(1450)$,  the approach also generates dynamically the 
 resonances $\sigma$ or $f_0(600)$, $\kappa$, $a_0(980)$ 
 and the main contribution to the $f_0(980)$.
 These resonances are 
generated even when no explicit resonances at tree level are included.
The basic point in UCHPT
 is to resum the right 
hand or unitarity cut to all orders, the source of the large corrections produced by the S-wave 
meson-meson interactions, and perform a chiral expansion of the rest, the so called 
interaction kernel \cite{kn,higs,nd}.
For the non-resonant isospins 3/2 and 2, not given in ref.\cite{nd},  the kernels 
correspond to the $\Opd$ CHPT amplitudes. In fig.\ref{fig:fitexp} we show the
reproduction of scattering data that is achieved by our S-wave amplitudes.
 
We only consider the S-waves since the lightest pseudoscalar self-energies are expected 
to be dominated by low energy physics where the P-waves are
kinematically suppressed by small factors of three-momentum squared. 
This is true even in the calculations of
loops since, after proper renormalization, the typical momentum in virtual
processes will be bounded and of the same order as the on-shell values.
Indeed, by a comparison of the order of magnitude of our results
with those obtained by regularizing
 the loops with a three-momentum cut-off, the latter must be around $0.3-0.4$~GeV.  
 At such energies only S-waves 
matters in very good approximation, which is also clear experimentally. 
Another signal of the suppression of P-wave dynamics in the self-energies is the fact that it 
only starts to contribute at the two chiral loop level.  In addition, we recall the discussion 
in section \ref{sec:intro}  concerning the large corrections to the chiral series typically 
induced by the resonant S-wave channels and the large ${\cal O}(p^6)$ counterterms calculated from
resonance exchanges in ref.\cite{mass}. 

\begin{figure}[ht]
\psfrag{P}{$P$}
\psfrag{Q}{$Q$}
\psfrag{Qb}{$\overline{Q}$}
\psfrag{Pb}{$\overline{P}$}
\psfrag{R}{$R$}
\psfrag{p}{p}
\psfrag{kk}{p-q}
\psfrag{k}{q}
\centerline{\epsfig{file=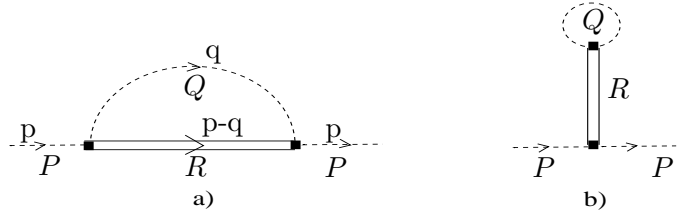,height=1.1in,width=3.5in,angle=0}}
\vspace{0.2cm}
\caption[pilf]{\protect \small
Fig.2a represents the $s-$channel exchange of scalar resonances $R$ 
in terms of the intermediate $Q$  and
initial $P$ pseudoscalars.
 A sum over all possible intermediate $Q$ and $R$ should be understood. 
 The fig.2b corresponds to the crossed exchange of the $I=0$ scalar resonances $R$.  
\label{fig:toy}}
\end{figure}

There is recently both experimental and theoretical overwhelming evidence of the existence of the
$I=0$ $\sigma$ \cite{iam,npa,nd,mixing,caprini,e791} and $I=1/2$ 
$\kappa$ \cite{iam,nd,mixing,kpi,e791,orsay} resonances. One also has the $I=0$ resonance 
$f_0(980)$ and the 
$I=1$ $a_0(980)$. These resonances have been considered quite often 
in the literature to form the
lightest nonet of scalar resonances \cite{eef,nd,mixing,pdg}, though there is not still consensus on
this issue. Since these resonances are the lightest scalar ones it is an interesting question to 
calculate their contributions to the pseudoscalar masses. In connection with this, 
  we plot in fig.\ref{fig:toy} those
contributions to the pseudoscalar self-energies that follow from the exchange of 
scalar resonances $R$ at the level of one loop with dressed propagators.\footnote{There 
 is a similar contribution to that of fig.\ref{fig:toy}b with the resonance leg 
 coupling directly to the vacuum without the tadpole 
in the upper extreme for the preexisting octet of scalar resonances at 1.4~GeV. 
These contributions were included e.g. in ref.\cite{kpi,jaminss} and 
their contributions to the chiral series for evaluating self-energies are negligible as they 
only contribute to $2L^r_8-L^r_5$ and $2L^r_6-L^r_4$, modulo small higher order contributions, with 
 a vanishing  result.}

\begin{figure}[ht]
\psfrag{P}{$P$}
\psfrag{Q}{$Q$}
\psfrag{Qb}{$\overline{Q}$}
\psfrag{Pb}{$\overline{P}$}
\centerline{\epsfig{file=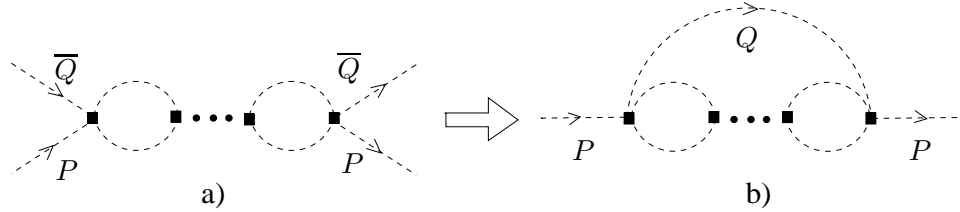,height=1.1in,width=5.in,angle=0}}
\vspace{0.2cm}
\caption[pilf]{\protect \small
Fig.3a represents the S-wave amplitude $P\overline{Q}\to P\overline{Q}$ and 
fig.3b is the diagram 
for the calculation of the self-energy of the pseudoscalar $P$ due to the intermediate 
pseudoscalar $Q$.
\label{fig:self}}
\end{figure}

\begin{figure}[ht]
\psfrag{P}{$P$}
\psfrag{Q}{$Q$}
\psfrag{Qb}{$\overline{Q}$}
\psfrag{Pb}{$\overline{P}$}
\psfrag{q}{$q$}
\centerline{\epsfig{file=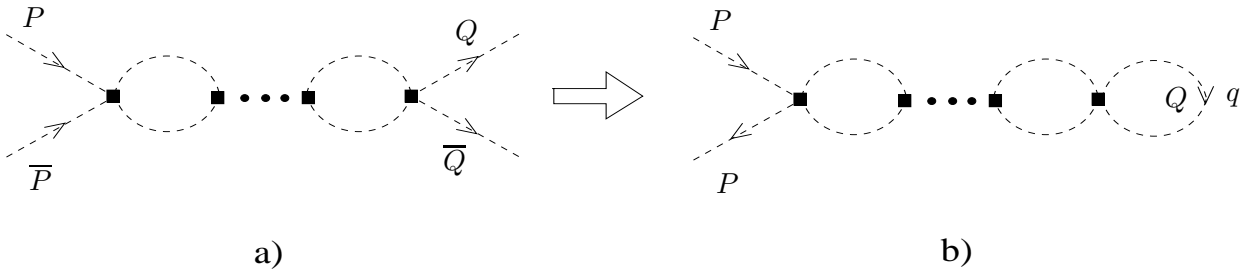,height=1.4in,width=6.5in,angle=0}}
\vspace{0.2cm}
\caption[pilf]{\protect \small
Tadpole diagram for the self-energy of the pseudoscalar $P$.
\label{fig:self3} }
\end{figure}

Within UCHPT \cite{nd} the resonances $f_0(980)$, $a_0(980)$, 
$\sigma$, $\kappa$ and an octet around 1.4~GeV appear as poles of the corresponding 
S-wave amplitudes once the right hand cut is resummed according to fig.\ref{fig:self}a. 
In this way fig.\ref{fig:toy}a is included in the evaluation of fig.\ref{fig:self}b and, analogously, 
fig.\ref{fig:toy}b is a contribution in fig.\ref{fig:self3}b. As remarked above,
 the lightest scalar resonances are generated through the
self-interactions of the pseudoscalars \cite{npa,nd}, with the kernel given by Chiral 
Symmetry \cite{wein,gl1,glsu3}. These scalar 
resonances in figs.\ref{fig:toy} better correspond to 
two pseudoscalars strongly correlated by the 
strong self-interactions.  Hence, in order to avoid double-counting one cannot talk independently
of the contributions of the 
unitarity loops and the light scalar resonances. Both must be calculated at once. This is why a
scheme like that in figs.\ref{fig:self} and \ref{fig:self3} is the appropriate one. 
 For the  resonances higher in mass, namely, the octet around 1.4~GeV, the situation is quite 
 different as they are preexisting resonances that are dressed by the interaction with the
  pseudoscalars. In this way,  tree level exchanges and pseudoscalar loops can be 
 differentiated in a perturbative way of thinking. 
 
 For the calculation of  
 fig.\ref{fig:self}b we proceed by analogy with  fig.\ref{fig:toy}a. Note that in 
 ref.\cite{Ddecays} it was shown that the pion and $K\pi$ scalar form factors are dominated by the
 Laurent poles of the $\sigma$ and $\kappa$ resonances,
  respectively,  for energies below the next resonance. In the case of the $I=0$ and 1/2 S-waves it
  was similarly shown that they are given by the previous poles plus a background, which mainly
  consists of just a constant which would give rise to a term analogous to fig.\ref{fig:toy}b 
  or \ref{fig:self3}b.
   It follows then that fig.\ref{fig:toy}a is a close analogy to 
  fig.\ref{fig:self}b.   
 After performing the integration over the null component of $q$ 
in the loop, see fig.\ref{fig:toy}a, one has two type of cuts. 
 The first  corresponds to having the pseudoscalar 
$Q$ on-shell, while for the second  the state on-shell is $\overline{R}$, running  
with opposite sense to that of $R$.  
The former involves the $P\overline{Q}$ scalar interaction mediated by the scalar resonance $R$.
 This is the
contribution we are seeking. The latter corresponds to the $P\overline{R}$ pseudoscalar interaction
mediated by the pseudoscalar $Q$ which is wildly off-shell and suppressed 
because the resonance energy is $-E_R$.
 
We then isolate from figs.\ref{fig:toy}a and \ref{fig:self}b
 the following scalar contribution to the self-energies of the pseudoscalars, 
\be
\Sigma^{U}_P=-\sum_Q \int\frac{d^3 k}{(2\pi)^3 2E_Q(\vk)} 
T_{P\overline{Q}\to P\overline{Q}}(s_1)~.
\label{self2}
\ee
with $Q=\{\pi^+,\pi^-,\pi^0,K^+,K^-,K^0,\overline{K}^0,\eta\}$ and 
$s_1=(M_P-E_Q(\vk))^2-\vk^2$ for $p=(M_P,\vec{0})$. 
The previous equation 
is invariant under Lorentz transformation as must be the case for a self-energy because 
a meson-meson scattering amplitude is a Lorentz scalar. 
 The sum  in eq.(\ref{self2}) is over all the 
species of pseudoscalars because  all of them give rise to different 
$P\overline{Q}_i\to P\overline{Q}_i$ amplitudes.

As stated above, one also has the corresponding diagrams figs.\ref{fig:toy}b and 
\ref{fig:self3}b.  Notice that in fig.\ref{fig:self} we were driven 
to consider the amplitudes $P\overline{Q}_i\to P\overline{Q}_i$, but  
one has to take into account the 
$t-$crossed process $P\overline{P}\to Q_i\overline{Q_i}$ as well. 
 We can evaluate the diagram 
in fig.\ref{fig:self3}b by performing the contour integration for $q^0$. Here one only 
has the singularity due to the $Q$ pseudoscalar and,
\be
\Sigma^{tad}_P=-\sum_{Q=\pi^0,\pi^+,K^+,K^0,\eta}\lambda_Q \int\frac{d^3 k}{(2\pi)^3 2E_Q(\vk)} 
T_{P\overline{P}\to Q\overline{Q}}(0)~.
\label{selftad}
\ee
In $\Sigma^{tad}_P$ one sums only over particles (without 
including the sum over antiparticles) since in the final state both particles and 
antiparticles happen and otherwise one is double-counting.  
Finally, $\lambda_Q=1$ for $\pi^+$, $K^+$ and $K^0$ and 
$\lambda_Q=1/2$ for $\pi^0$ and $\eta$, because they are their own antiparticles. These 
warnings also  happen in the standard calculations of tadpoles with the 
use of Feynman rules because of the same reasons. 
The self-energy of the pseudoscalar $P$ is then given by the sum,

\be
\Sigma_P=\Sigma^{U}_P+\Sigma_P^{tad}~,
\label{sigtot}
\ee
 In terms of the amplitudes for the different channels
we have for the $\pi$, $K$ and $\eta$ self-energies,
\ba
\Sigma_\pi&=&-\int\frac{d^3 k}{(2\pi)^3 2
E_Q(\vk)}  \left[\frac{10}{3} T^{I=2}_{\pi\pi\to \pi\pi}(s)
+\frac{2}{3}T^{I=0}_{\pi\pi\to \pi\pi}(s)
+\frac{8}{3}T^{I=3/2}_{\pi\bar K\to \pi\bar K}(s)
+\frac{4}{3}T^{I=1/2}_{\pi\bar K\to \pi\bar K}(s)
+T^{I=1}_{\pi\eta\to \pi\eta}(s) \right. \nn\\
\qquad &+&\left. T^{I=0}_{\pi\pi\to \pi\pi}(0)
+\frac{2}{\sqrt{3}}T^{I=0}_{\pi\pi\to K\bar K}(0)
-\frac{1}{\sqrt{3}}T^{I=0}_{\pi\pi\to \eta\eta}(0)\right]\nn \\
\Sigma_K&=&-\int\frac{d^3 k}{(2\pi)^3 2
E_Q(\vk)}  \left[
2 T^{I=3/2}_{\pi K\to \pi K}(s)
+T^{I=1/2}_{\pi K\to \pi K}(s)
+\frac{3}{2}T^{I=1}_{K\bar K\to K\bar K}(s)
+\frac{1}{2}T^{I=0}_{K\bar K\to K\bar K}(s) \right. \nn \\
 &+&\left. \frac{3}{2}T^{I=1}_{K K\to K K}(s)
+\frac{1}{2}T^{I=0}_{K K\to K K}(s)
+T^{I=1/2}_{K\eta\to K\eta}(s) \right. \nn\\
 &+&\left. \frac{\sqrt{3}}{2}T^{I=0}_{K\bar K\to \pi\pi}(0)
+T^{I=0}_{K\bar K\to K\bar K}(0)
-\frac{1}{2}T^{I=0}_{K \bar K\to \eta\eta}(0)\right]\nn \\
\Sigma_\eta&=&-\int\frac{d^3 k}{(2\pi)^3 2
E_Q(\vk)}  \left[
3 T^{I=1}_{\pi \eta\to \pi \eta}(s)
+4 T^{I=1/2}_{\eta K\to \eta K}(s)
+2T^{I=0}_{\eta\eta\to\eta\eta}(s) \right. \nn \\
 &-&\left.\sqrt{3} T^{I=0}_{\eta\eta\to \pi\pi}(0)
-2T^{I=0}_{\eta\eta\to K\bar K}(0)
+T^{I=0}_{\eta\eta\to \eta\eta}(0)\right]
\label{finaleq}
\ea

Notice that we are employing just the S-wave contribution to the full strong amplitude 
$T_{ij}$, and this is why we have kept only the $s$-Mandelstam variable in the argument 
of $T_{ij}(s_1)$, dropping the $t$ dependence in eq.(\ref{finaleq}). This is what automatically 
occurs when considering the intermediate scalar resonances in figs.\ref{fig:toy}a, b.
 The Clebsch-Gordan coefficients in front of the S-wave amplitudes in the previous equations are obtained
 from a standard isospin analysis. As an example,
  for the $\pi^+$ self-energy one has the S-wave amplitudes,
 \ba
 T_{\pi^+\pi^-\to \pi^+\pi^-}+T_{\pi^+\pi^+\to\pi^+\pi^+}+T_{\pi^+\pi^0\to\pi^+\pi^0}&=&
\left(\frac{1}{6}T^{I=2}_{\pi\pi\to\pi\pi}+\frac{1}{3}T^{I=0}_{\pi\pi\to\pi\pi}\right)+T^{I=2}_{\pi\pi\to\pi\pi}
+\frac{1}{2}T^{I=2}_{\pi\pi\to\pi\pi}\nn\\
&=&\frac{5}{3}T^{I=2}_{\pi\pi\to\pi\pi}+\frac{1}{3}T^{I=0}_{\pi\pi\to\pi\pi}~.
 \ea
 To finally obtain the coefficients in $\Sigma_\pi$ one has to multiply by 2 the $T^I_{\pi\pi\to\pi\pi}$
 S-waves in the last expression because in refs.\cite{npa,nd} one uses 
 the $I=0$, 2, $\pi\pi$ states (and also the $I=0$
 $\eta\eta$ one) normalized to 1/2. In this way, these symmetric states under the exchange of the two
 pions can be treated as if the latter were distinguishable. For more details  see
 refs.\cite{npa,nd}.

The integral in eq.(\ref{self2}) for $\Sigma^{U}_P$ is ultraviolet divergent
 since for large three-momentum 
the measure grows as $\vk^2$ and $T_{P\overline{Q}\to P\overline{Q}}(s_1)$ tends to $1/\log s_1$ 
or constant for $|s_1|\to \infty$. This, together
 with the factor $E_Q(\vk)$ in the denominator, gives rise to a quadratic divergence. 
For $\Sigma^{tad}_P$, eq.(\ref{selftad}), one has the same type of divergence as  
$T_{P\overline{P}\to Q \overline{Q}}$ is evaluated at $s_1=0$ and it is a constant.

\begin{figure}[H]
\psfrag{P}{$P$}
\psfrag{Q}{$Q$}
\psfrag{Qb}{$\overline{Q}$}
\psfrag{Pb}{$\overline{P}$}
\centerline{\epsfig{file=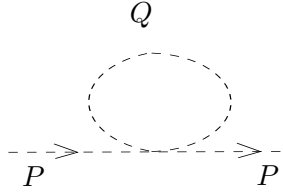,height=1.in,width=1.5in,angle=0}}
\vspace{0.2cm}
\caption[pilf]{\protect \small
The one loop  contribution of figs.\ref{fig:self}b and \ref{fig:self3}b at ${\Opc}$.
  This is the only loop appearing at $\Opc$ CHPT.
\label{fig:selfp4}}
\end{figure} 

A general coupled channel S-wave T-matrix  has the structure  \cite{nd},
\be
T=\left[I+{\cal K}\cdot G\right]^{-1}\cdot {\cal K}~,
\label{tmat}
\ee
with ${\cal K}$ the interaction kernel
 and $G$ corresponds to the unitarity bubbles shown in fig.\ref{fig:self}. 
For more details see ref.\cite{nd}.  
The previous equation has the expansion,
\be
T=\left[I-{\cal K}\cdot G+{\cal K}\cdot G\cdot {\cal K}+\ldots\right]\cdot {\cal K}~.
\label{bfdouble}
\ee
We notice that when $T={\cal K}_2$ (the lowest order CHPT amplitudes),
 one has  no unitarity bubbles in fig.\ref{fig:self}b and fig.\ref{fig:self3}b,
and then both figures  give the same diagram. This only happens for 
${\cal K}_2$ at lowest order in CHPT. The local term ${\cal K}_2$ gives 
rise to the one loop diagram shown in fig.\ref{fig:selfp4}, the only one that appears 
at next-to-leading order (NLO) 
in CHPT (at this order the rest of contributions come from local counterterms \cite{gl1,glsu3}).
 In order 
to avoid this double-counting we then subtract the contribution in fig.\ref{fig:selfp4} to the 
 integral in eq.(\ref{self2}), fig.\ref{fig:self}b. This is done explicitly below 
in eq.(\ref{extv2}).

\subsection{Regularization and renormalization}

We now proceed with the evaluation of the integral in eq.(\ref{self2}) for $\Sigma^{U}_P$.
 The integral 
in eq.(\ref{selftad}) for $\Sigma^{tad}_P$ is a particular case of the previous one. 
Firstly, we perform the angular integration and change to the energy as the integration variable.
\be
\int \frac{d^3 k}{(2\pi)^3 2E_Q(\vk)}T_{P\overline{Q}\to P\overline{Q}}(s_1)=
\int_{M_Q}^\infty \frac{|\vk| dE_Q}{4\pi^2} T_{P\overline{Q}\to P\overline{Q}}(s_1)~.
\label{intdk}
\ee
 We now introduce the new variable
\be 
t=M_Q/E_Q~.
\label{deft}
\ee
 As the resulting integral  diverges quadratically 
for $t\to 0$, we change the lower limit to 
 $\varepsilon$, positive and small, taking at the end the limit $\varepsilon\to 0^+$. 
 \be
\frac{M_Q^2}{(2\pi)^2}\int_\varepsilon^1 dt\frac{ \sqrt{1-t^2}}{t^3} T_{P\overline{Q}\to P\overline{Q}}(s_1)~,
\label{epsilon}
\ee
with $s_1=M_P^2+M_Q^2-2 M_P M_Q/t$.
Notice that since $\epe=M_Q/\Lambda$, with $\Lambda\gg M_Q$ 
being the upper limit 
in the three-momentum  of eq.(\ref{intdk}), the previous limit is equivalent to 
$\Lambda\to +\infty$. In performing this limit
we must identify those terms that scales as $1/\varepsilon^2$, $1/\varepsilon$ (equivalently 
as $\Lambda^2$, $\Lambda$) and remove them from the result.
In order to remove infinities one must work out algebraic expressions for the previous integral that 
explicitly show these diverging powers of $\epe$. The problem of accomplishing 
 this aim is the complicated matrix expression for the strong amplitude that 
prevents us from giving a close expression for the integral in eq.(\ref{epsilon}).
 However, we are interested in isolating algebraically the dependence in 
$\varepsilon$, and this can be done by performing a dispersion relation representation for  
$T_{P\overline{Q}\to P\overline{Q}}(s_1)$ in the physical Riemann sheet. 
 
  The $T_{P\overline{Q}\to P\overline{Q}}(s_1)$ amplitude from 
ref.\cite{nd} 
has only one cut, the right or unitarity one, and tends to $1/\log s_1$ or constant when 
$s_1\to \infty$. In order to perform the dispersion relation we take as integration
 contour a circle 
in the infinity deformed to engulf the real axis along the right hand cut, 
$s_{th}<s_1<\infty$, with $s_{th}$ the lightest meson-meson threshold with the 
$P\overline{Q}$ S-wave quantum numbers. 
We take one subtraction because of the aforementioned behaviour
 of $T_{ij}(s_1)$ at infinity. We then have:
\be
T_{ij}(s_1)=T_{ij}(s_2)+\frac{s_1-s_2}{\pi}\int_{s_{th}}^\infty
ds'\frac{\hbox{Im}T_{ij}(s')}{(s'-s_1)(s'-s_2)}+\sum_{\ell=1}^N
\frac{(s_2-s_1)R_{ij}^{(\ell)}}{(s^R_\ell-s_1)(s^R_\ell-s_2)}~,
\label{disp}
\ee
with $s_2$  the point where the subtraction has been performed. In addition, 
$N$ is the number of poles of $T_{ij}(s)$ in the physical Riemann sheet 
and $R_{ij}^{(\ell)}$ is the residue
 of the $\ell_{th}$ pole at the position $s^R_\ell$.  Because of the Schwartz 
reflection principle any partial wave satisfies that $T_{ij}(s^*)=T_{ij}(s)^*$, so that poles 
always appear as pairs in relative complex conjugate positions with complex conjugate residua. 
Taking then $s_2$ real in eq.(\ref{disp}) the contribution from the sum of poles 
is real for $s_1$ along the real $s$-axis. 

 A physical partial wave should not have any pole on the physical sheet with non-vanishing 
 imaginary part. But notice that the lowest order CHPT S-wave amplitudes 
 in $I=1/2$ ($K \pi\to K\pi$, $K\pi\to K\eta$ and $K\eta \to K\eta$) have each of them 
 one pole at $s=0$ due to the kinematical reason of having different masses, see ref.\cite{nd} 
 for explicit expressions of those amplitudes. This pole at $s=0$ for the interaction kernel 
drives the appearance of a pole in the final $K\pi\to K\pi$ S-wave for a 
real $s$ value, typically at around $s=-0.25$~GeV$^2$,  not far away from its lowest order value. This poles  
gives rise to a non-negligible effect in the calculation of the $\pi$ self-energy. 
There are also other poles with non-zero imaginary part.  Reassuringly, 
these poles are always far away of the physical axis and their contributions 
to the pseudoscalar self-energies are negligible, as we have numerically checked. 

Let us proceed with the evaluation of the renormalized value of the 
divergent integral eq.(\ref{epsilon}) making use of the
representation given in eq.(\ref{disp}) for the S-wave T-matrix elements.
 Both the contributions from the unitarity cut and poles involve 
the same sort of integral,
\be
J(s',s_2)=\int_\varepsilon^1dt \frac{\sqrt{1-t^2}}{t^3}\frac{s_1(t)-s_2}{s'-s_1(t)}~.
\label{jsps2}
\ee
For the sum over the poles in eq.(\ref{disp}) the variable 
 $s'$ must be replaced by $s^R_{\ell}$. 
 For convenience, new variables $a$ and $b$ are introduced,  
 \ba
 a=\nu-\frac{s_2}{2 \mmp \mmq}~,~ b=\nu-\frac{s'}{2 \mmp \mmq}~,
\label{defab}
 \ea
with $\nu=(\mmp^2+\mmq^2)/(2\mmp\mmq)$. Notice that one 
can also write $1/t=\nu-s_1/2M_P M_Q$. Then,
\ba
J(s',s_2)&=&-\int_\epe^1 dt \frac{\sqrt{1-t^2}}{t^3}\frac{1-at}{1-bt}=-\int_\epe^1 \frac{\sqrt{1-t^2}}{t^3}+
(a-b)\int_\epe^1 dt \frac{\sqrt{1-t^2}}{t^2}\nn\\
&+&(a-b)b\int_\epe^1 dt\frac{\sqrt{1-t^2}}{t}-(a-b)b\int_\epe^1 
dt \frac{\sqrt{1-t^2}}{t-1/b}~.
\label{int27}
\ea
The first integral is elementary, although its renormalized value requires some comments. 
\be
-\int_\epe^1 dt \frac{\sqrt{1-t^2}}{t^3}=-\frac{1}{2}\frac{\sqrt{1-\epe^2}}{\epe^2}+
\frac{1}{2}\log\frac{1+\sqrt{1-\epe^2}}{\epe}~.
\label{firstin}
\ee
Now, when sending $\epe\to 0^+$ both terms on the r.h.s.  diverge. Performing 
a power series around $\epe=0$ one has $-1/2\epe^2+(1-2\log \epe/2)/4+{\cal O}(\epe^2)$. The first 
term in this sum diverges like $\epe^{-2}$ and should
be reabsorbed by appropriate chiral counterterms. The terms with positive powers of $\epe$, not explicitly shown, 
vanish for $\epe\to 0$. Then, the remaining contribution is,
\be
\frac{1}{4}\left(1-2\log \frac{M_Q}{2\Lambda}\right)=\frac{1}{4}\left(1-2\log
\frac{M_Q}{2\mu}\right)-\frac{1}{2}\log\frac{\mu}{\Lambda}~,
\label{rmu}
\ee 
where the renormalization scale $\mu$ is introduced. 
 In sending $\Lambda \to +\infty$ ($\epe\to 0^+$), again 
the term $\log \mu/\Lambda$ diverges and should be reabsorbed by the appropriate counterterms. Then 
one has:
\be
-\int_0^1 dt \frac{\sqrt{1-t^2}}{t^3} \doteq \frac{1}{4}\left(1-2\log
\frac{M_Q}{2\mu}\right)~,
\label{onehas}
\ee
where the symbol $\doteq$, used for expressing our renormalized integrals, means equal up to 
terms divergent as $\epe^{-2}$, $\epe^{-1}$ or $\log \epe$, when $\epe \to 0^+$. 
 When giving our results
we shall vary $\mu$ between $0.5$ and $1.2$~GeV, providing in this way 
an estimate of possible constants 
of order one that may change depending on the  renormalization scheme chosen.
 E.g. the difference between our
procedure and the results in the modified $\overline{MS}$ scheme in dimensional regularization 
used in CHPT \cite{gl1,glsu3}. 
 This variation in $\mu$ will be a source of uncertainty
  in our results and it will be consider in the error analyses. In ref.\cite{iam} the unitarity
  bubble $G$ in eq.(\ref{tmat}) was also calculated within a  cut-off scheme. If one sends this cut-off to
  infinity only a logarithmic divergence remains, as we have here. By varying $\mu$ within
  the range mentioned above, one is accounting for finite constant terms, like the subtraction constant 
used in $G$.
 
 The second and third integrals on the right hand side of eq.(\ref{int27}) can be treated similarly. 
 For the last one, 
 \be
\int dt\frac{\sqrt{1-t^2}}{t-c}=\sqrt{1-t^2}-c\,\arcsin t+(1-c^2)\int\frac{dt}{(t-c)\sqrt{1-t^2}}~,\nn\\
\ee
with $c=1/b$ and
\ba
\widetilde{I}(b)=\int_\epe^1\frac{dt}{(t-1/b)\sqrt{1-t^2}}&=&
i)~ \frac{1}{\sqrt{1-1/b^2}}\log\frac{1+\sqrt{1-1/b^2}}{\epe+1/b}~,~ b>1~,\nn\\
&& ii)~ \frac{1}{\sqrt{1-1/b^2}}\log\frac{1+\sqrt{1-1/b^2}}{\epe-1/b}~,~ b<-1~,\nn\\
&& iii)~\frac{-1}{\sqrt{1/b^2-1}}\left(\frac{\pi}{2}+\arcsin b\right)~,~0<b<1~,\nn\\
&& iv)~\frac{1}{\sqrt{1/b^2-1}}\left(\frac{\pi}{2}+\arcsin b\right)~,~-1<b<0~,
\label{wdi}
\ea

Adding all the integrals that contribute  in eq.(\ref{int27}),
\be
J(s',s_2)=\frac{1}{4}+\frac{1}{2}\log\frac{2}{\epe}+(a-b)b\log\frac{2}{\epe}-(a-b)b(1-1/b^2)\,\widetilde{I}(b)~.
\ee
Now, in the expressions given for $\widetilde{I}$ with $|b|>1$ in eq.(\ref{wdi}), the first two lines, 
one has $\log (\epe+1/|b|)=\log \epe+\log(1+1/\epe |b|)$. This 
 last term gives rise to an infinite
series of divergent terms in powers of $1/(|b|\,\epe)^n$  for $|b|>1$, equivalently in powers of $(\Lambda/b)^n$,
which are removed in the regularization process. Substituting also
 $\log \epe$ by $\log \mmq/\mu$, 
 introducing the renormalization scale $\mu$, as explained before, then
\ba
\label{jsex}
&&J(s',s_2)\doteq\nn\\
i)&& J_+ (a,b)=\frac{1}{4}+\frac{1}{2}\log\frac{2\mu}{\mmq}+(a-b)b\log2
-(a-b)b(1-\sqrt{1-1/b^2})\log\frac{\mmq}{\mu}
\nn\\
&&-(a-b)b\sqrt{1-1/b^2}\log(1+\sqrt{1-1/b^2})~, ~|b|>1~,\\
ii)&& J_-(a,b)=\frac{1}{4}+\frac{1}{2}\log\frac{2 \mu}{M_Q}
+(a-b)b\log\frac{2\mu}{\mmq}-(a-b)\sqrt{1-b^2}\left(\frac{\pi}{2}+\arcsin b\right)~,~|b|<1~.\nn
\ea

 The dispersive integral contribution to the self-energy of the pseudoscalar $P$ 
 by the intermediate $Q$ is:
 \ba
\label{intdis}
&&I_{disp}^{PQ}=\int_\varepsilon^1 dt\frac{\sqrt{1-t^3}}{t^3}\frac{s_1-s_2}{\pi}\int_{s_{th}}^\infty ds'
\frac{\hbox{Im}T_{ij}(s')}{(s'-s(t))(s'-s_2)}\doteq\nn\\
&&i) \hbox{~If~}(\mmp-\mmq)^2>s_{th}\nn\\
&&\frac{1}{\pi}\int_{s_{th}}^{(M_P-M_Q)^2} ds'\frac{\hbox{Im}T_{ij}(s')}{(s'-s_2)}J_+(s',s_2)+
\frac{1}{\pi}\int_{(M_P-M_Q)^2}^{(\mmp+\mmq)^2}ds'\frac{\hbox{Im}T_{ij}(s')}{(s'-s_2)}J_-(s',s_2)\nn\\
&&+
\frac{1}{\pi}\int_{(\mmp+\mmq)^2}^{+\infty} ds'\frac{\hbox{Im}T_{ij}(s')}{(s'-s_2)}J_+(s',s_2)\nn\\
&&ii) \hbox{~If~}(\mmp-\mmq)^2<s_{th}\nn\\
&&\frac{1}{\pi}\int_{s_{th}}^{(M_P+M_Q)^2} ds'\frac{\hbox{Im}T_{ij}(s')}{(s'-s_2)}J_-(s',s_2)+
\frac{1}{\pi}\int_{(M_P+M_Q)^2}^{+\infty}ds'\frac{\hbox{Im}T_{ij}(s')}{(s'-s_2)}J_+(s',s_2)
\ea
Notice that $(\mmp+\mmq)^2\geq s_{th}$. 
Let us remark that $J_+(s',s_2)$ vanishes like $-1/s'$ for $s'\to\infty$, 
 as it is clear from eq.(\ref{jsex}). This is why
all the last integrals in the previous equations converge for $s'\to +\infty$.

 The subtraction constant in eq.(\ref{disp}), $T_{ij}(s_2)$, contributes as,
\be
I_{subs}^{PQ}=T_{ij}(s_2)\int_\epe^1dt \frac{\sqrt{1-t^2}}{t^3}\doteq
-T_{ij}(s_2)\frac{1}{4}\left(1-2\log\frac{\mmq}{2\mu}\right)~,
\label{intsubs}
\ee
from eq.(\ref{onehas}).   The contribution of the sum of poles in eq.(\ref{disp}) is:
\be
I_{pole}^{PQ}=-\sum_{\ell=1}^N
\frac{R_{ij}^{(\ell)}}{(s^R_\ell-s_2)}J_+(a,b(s^R_\ell))~.
\label{intpolsum}
\ee
We have only $J_+(a,b)$ since for this case always $|b|>1$ as the poles happen far
away in the complex plane or on the negative $s$-axis. 
 If  $s_2$ is taken such that $a=0$ then $J(a,b)$ will vanish like ${\cal O}(b^{-2})$ 
 for $|b|\to\infty$, as follows from eq.(\ref{jsex}). One can take advantage of this fact  
 since  the pole contributions will tend to
vanish because  $s_\ell^R$ and the resulting $|b|$ are large.
 Thus,  we fix $s_2=s_A=\mmp^2+\mmq^2$ in the following so that $a=0$,  
 eq.(\ref{defab}). This choice also makes that all the contributions in 
our final equation for the self-energies, eq.(\ref{selff}) below, have natural size.

One still has  to remove the $\Opc$ contribution
 to $I_{disp}^{PQ}$ in order to avoid double-counting with the 
tadpole contribution, as 
explained above after eq.(\ref{bfdouble}). We then evaluate eq.(\ref{self2}) 
with $T_{P\overline{Q}\to P\overline{Q}}$ given by its $\Opd$ expression. A general 
CHPT S-wave amplitude at $\Opd$ has the form,
\be
T_{ij}^{(2)}(s)=A+B s +\frac{C}{s}~.
\label{opdchpt}
\ee
The last term in the sum only appears for the $I=1/2$ $K\pi$ and $K\eta$ coupled channel amplitudes
involving two particles with different masses. 
Then,
\ba
&&\int_0^1 dt\frac{\sqrt{1-t^2}}{t^3}\left(A+B s_1(t)+\frac{C}{s_1(t)}\right)\nn\\
&&=
\int_0^1 \frac{\sqrt{1-t^2}}{t^3}\left(A+ s_A B\right)-2M_P M_Q B\int_0^1 dt \frac{\sqrt{1-t^2}}{t^4}
+
\frac{C}{s_A}\int_0^1 dt \frac{\sqrt{1-t^2}}{t^2}\frac{1}{t-f}~,
\ea
with $f=1/\nu$ ($f<1$ when $C\neq 0$)  and $s_A=M_P^2+M_Q^2$. 
The renormalized second integral in this equation vanishes as can easily worked out. 
We then have, 
\ba 
I_{rem}^{PQ}&=&T_{ij}^{(2)}(s_A)\int_0^1 dt\frac{\sqrt{1-t^2}}{t^3}
+\frac{C}{s_A}\int_0^1 dt \frac{\sqrt{1-t^2}}{t^3}
\frac{f}{t-f}\nn\\
&=&-T_{ij}^{(2)}(s_A)\frac{1}{4}(1-2\log \frac{M_Q}{2\mu})+\frac{C}{s_A}
 J_+(0,1/f)~.
\label{extv2}
\ea

The other contribution to $\Sigma_P$ is $\Sigma_P^{tad}$, eq.(\ref{sigtot}). 
Its calculation, eq.(\ref{firstin}), is straightforward  and gives,
\be
I_{tad}^{PQ}=\lambda_{Q}T_{ij}(0)\int_\epe^1dt \frac{\sqrt{1-t^2}}{t^3}\doteq
-\lambda_Q\frac{T_{ij}(0)}{4}\left(1-2\log\frac{\mmq}{2\mu}\right)~.
\label{inttad}
\ee
 Hence, for eq.(\ref{sigtot}) one has:
 \be
\Sigma_P=-\sum_Q \frac{\mmq^2}{(2\pi)^2}\left(I_{subs}^{PQ}+I_{disp}^{PQ}+I_{pole}^{PQ}-I_{rem}^{PQ}\right)
-\sum_{Q'} \frac{M_{Q'}^2}{(2\pi)^2} I_{tad}^{PQ'}
\label{selff}
 \ee
with $Q=\{\pi^0,\pi^+,\pi^-,K^0,\overline{K}^0,K^+,K^-,\eta\}$ and 
$Q'=\left\{\pi^0,\pi^+,K^+,K^0,\eta\right\}$. Notice that after the 
subtraction of $\sum_Q I_{rem}^{PQ} \,M_Q^2/(2\pi)^2$ to the previous equation,  
the only contribution at $\Opc$ of our result comes from
$I_{tad}^{PQ} M_Q^2/(2\pi)^2$ when $T_{ij}(0)$ in eq.(\ref{inttad}) 
is substituted by $T_{ij}^{(2)}(0)$, the corresponding 
lowest order CHPT amplitude. It is important to remark that for this $\Opc$ case 
we have checked that  $\sum_{Q'} I^{PQ'}_{tad}\,M_{Q'}^2/(2\pi)^2$ 
reproduces the CHPT infrared logarithms 
on the quark masses in CHPT at ${\Opc}$ \cite{glsu3}. 
This is one among the many explicit calculations 
 in the literature showing that by sending $\Lambda\to +\infty$
  and removing the power divergences (which necessarily disappears in dimensional regularization)
 one recovers the dimensional regularization results with the differences 
 reabsorbed in the chiral counterterms from the tree level contributions \cite{hem,epe,bura}.

\subsection{Mass equations}

We  often remove the $\Opc$ contribution to $\Sigma_P$ and denote by
 $\Sigma_P^H$ the ${\cal O}(p^6)$ and higher order contributions. 
 The latter is obtained with the replacement of $T_{ij}(0)$ by  
 $T_{ij}(0)-T_{ij}^{(2)}(0)$ in eq.(\ref{inttad}) for the tadpole contribution. 
  The   $\Opc$ CHPT expressions for the pseudoscalar masses are, ref.\cite{glsu3}:
\ba
\label{chptmasses}
M_\pi^2&=&\krig{M}_\pi^2\left\{1+\mu_\pi-\frac{1}{3}\mu_\eta+2\hat{m}K_3+K_4\right\}~,\nn\\
M_K^2&=&\krig{M}_K^2\left\{1+\frac{2}{3}\mu_\eta+(\hat{m}+m_s)K_3+K_4\right\}~,\\
M_\eta^2&=&\krig{M}_\eta^2\left\{1+2\mu_K-\frac{4}{3}\mu_\eta+\frac{2}{3}(\hat{m}+m_s)K_3+
K_4\right\}+\krig{M}_\pi^2\left\{-\mu_\pi+\frac{2}{3}\mu_K+\frac{1}{3}\mu_\eta\right\}+K_5~,\nn
\ea
where,
\ba
\label{kchpt}
K_3&=&\frac{8B_0}{F_0^2}(2L_8^r-L_5^r)~,~K_4=(2\hat{m}+m_s)\frac{16 B_0}{F_0^2}(2L_6^r-L_4^r)~,\nn\\
K_5&=&(m_s-\hat{m})^2\frac{128}{9}\frac{B_0^2}{F_0^2}(3L_7+L_8^r)~,~
\mu_Q=\frac{1}{32\pi^2}\frac{\krig{M}_P^2}{F_0^2}\log\frac{\krig{M}_P^2}{\mu^2}~.
\ea
The symbol $\krig{M}_P$ denotes the bare mass of the pseudoscalar $P$ corresponding to the lowest order 
CHPT result,
\be
\krig{M}_\pi^2=2B_0\hat{m}~,~~\krig{M}_K^2=B_0(\hat{m}+m_s)~,~~\krig{M}_\eta^2=\frac{2}{3}B_0(\hat{m}+2m_s)~,
\label{b0mq}
\ee
and $B_0$ measures the strength of the quark condensate $\la 0|\bar{q}q|0\ra=-F_0^2 B_0+{\cal O}(m_q^2)$
 in the SU(3) chiral limit.
 We will denote the $\Opc$ CHPT self-energies by $\Sigma_P^{4\chi}=M_P^2-\krig{M}_P^2$, 
 eq.(\ref{chptmasses}). Notice that from  eq.(\ref{b0mq}) $\krig{M}_\eta^2$ always satisfies the 
Gell-Mann-Okubo mass relation,
\be
\krig{M}_\eta^2=\frac{1}{3}(4\mkk^2-\mkp^2)~.
\label{okubo}
\ee
 On the other hand, the $L_i^r$ coefficients are the renormalized values 
of the low energy CHPT counterterms at $\Opc$ from ref.\cite{glsu3} within the $\overline{MS}-1$ scheme. 

The mass equation  one has to solve is,
\be
M_P^2=\krig{M}_P^2+\Sigma_P(\krig{M}_Q^2;M_Q^2)~,
\label{eq1}
\ee
with $\Sigma_P$ given in eq.(\ref{selff}). The dependence 
on the bare masses $\krig{M}_Q^2$ originates from the one of  
 the interaction kernel ${\cal K}$ \cite{nd}. 

If we impose that our results  incorporate the $\Opc$ CHPT self-energies 
and resum higher order contributions, the following equations arise,
\be
M_P^2=\krig{M}_P^2+\Sigma_P^{4\chi}(\krig{M}_Q^2;L_i^r)+\Sigma_P^H(\krig{M}_Q^2;M_Q^2)~.
\label{eq2}
\ee

Our analysis is performed in the isospin limit, $m_u=m_d=\hat{m}$ and no electromagnetic 
contributions. The symbols $M_P^2$ above refer  
to the  
masses of the lightest pseudoscalars in the isospin limit. To determine these masses 
we employ the Dashen 
theorem \cite{dashen} which states that at the lowest nontrivial order in $e^2$
  the neutral particles $\pi^0$ and $K^0$ do not receive 
electromagnetic contributions, while these are the same for $M_{\pi^+}^2$ and $M_{K^+}^2$. In this way, 
\be
M_\pi^2=M_{\pi^0}^2\simeq (135 \hbox{ MeV})^2 ~,
~M_K^2=(M_{K^+}+M_{K^0}^2-M_{\pi^+}^2+M_{\pi^0}^2)/2\simeq 
(495 \hbox{ MeV})^2~,
\label{massisos}
\ee
are taken as the pion and kaon masses in the isospin limit, respectively. 
Violations of the previous relation for the pion mass are expected to be tiny,
${\cal O}((m_u-m_d)^2)$ \cite{glsu3}, from the pure QCD side, and ${\cal O}(e^2 m_\pi^2/(8\pi^2 f_\pi^2))$ 
from electromagnetic interactions for the $\pi^0$ mass. 
More substantial seems to be the violations of the Dashen theorem for the kaon masses, 
 of order $e^2 M_K^2$ \cite{22,23,24}.
 So that, if we write $M_K^2=\left[M_{K^+}^2+M_{K^0}^2-
(1+\delta D)(M_{\pi^+}^2-M_{\pi^0}^2)\right]/2$, with $\delta D$ measuring the deviation 
with respect to 
the Dashen theorem, then as a conservative estimate, 
$\delta D$ is expected to be in the range $0-2$ \cite{milc} 
from refs.\cite{23,24}. But even so, the changes in $M_K^2$ are at most a 
$0.5$\%, which can be 
discarded within the definitely larger  uncertainties of our calculations, to be shown in the 
next sections. 


\section{Dynamical self-energies}
\label{sec:dyn}
We restrict in this section to eq.(\ref{eq1}) and solve
 for $\krig{M}_{\pi,K}^2$ (equivalently, for $B_0 \hat{m}$ and $B_0 m_s$) 
 and for $M_\eta^2$. In this section the $\eta$ is actually the $\eta_8$ since 
we are not including any source of $\eta-\eta'$ mixing. 
The distinction between the $\eta$ and $\eta_8$ is accomplished
 in the next section. 

 In perturbative studies one 
solves iteratively eq.(\ref{eq1}). In this way, for the first round, 
 we will use physical masses 
  in all the arguments of the self-energies 
  $\Sigma_P(\krig{M}_Q^2;M_Q^2)$ with the resulting values
\be
\mkp=108\pm 4~,~\mkk= 422\pm 18~,~M_{\eta_8}=660\pm 17~\hbox{MeV}.
\label{valnox1}
\ee
  The errors are calculated by performing a Monte-Carlo sampling
  of the free parameters of the T-matrix of ref.\cite{nd}\footnote{ We have made a new fit, including some more recent data, and the resulting fit has central values for the free parameters that 
are  compatible with those of ref.\cite{nd}, within the shown errors in this reference.} and 
varying the renormalization scale $\mu$ introduced above in the range between $0.5$ and $1.2$~GeV.
 This will
be the standard procedure to evaluate errors in our work and should be understood in the following. 
Performing one more iteration in eq.(\ref{eq1}), i.e. 
using the previous bare masses in the appropriate arguments of the
 self-energies,
it results
\be
\mkp=158\pm 7~,~\mkk= 511\pm 12~,~M_{\eta_8}=627\pm 15~\hbox{MeV}.
\label{valnox2}
\ee
 
We observe significant  differences between eqs.(\ref{valnox1}) and (\ref{valnox2}) for 
$\mkp$ and $\mkk$,
 which is a clear indication that higher orders corrections, 
 of ${\cal O}(p^6)$ and superior, are relevant for the calculation of the lightest
  pseudoscalar self-energies in SU(3). This was also observed in 
   refs.\cite{bijtal,mass}, where the masses were calculated at ${\cal O}(p^6)$ in CHPT.

Because of the large variation between the values 
in eqs.(\ref{valnox1}) and (\ref{valnox2}), 
 eq.(\ref{eq1}) should be solved exactly. 
 This is done by minimizing the function,
\ba
S(\mkp,\mkk)&=&\left(\frac{\mkp^2+\Sigma_\pi(\krig{M}_Q^2;M_Q^2)}{M_\pi^2}-1\right)^2
+\left(\frac{\mkk^2+\Sigma_K(\krig{M}_Q^2;M_Q^2)}{M_K^2}-1\right)^2\nn\\
&+&\left(\frac{\mke^2+\Sigma_\eta(\krig{M}_Q^2;M_Q^2)-M_{\eta_8}^2}
{M_K^2}\right)^2~.
\label{smi}
\ea

This method is more robust than the iterative one and 
 it can provide solutions even when  the later  
does not converge. On the other hand, 
since we are now varying the bare masses employed in the kernels for the
T-matrices, we refit the scattering data in fig.\ref{fig:fitexp} 
for every value of the given bare masses. Then, the T-matrix changes
as a function of the values  
employed in each step.  The following results are obtained,
\be
\mkp=126\pm 4~,~\mkk=476\pm 10~,~M_{\eta_8}= 635 \pm 15~\hbox{MeV}.
\label{valnox3}
\ee

One observes significant variations between the exact result and the first and second iterated solutions, 
eq.(\ref{valnox1}) and (\ref{valnox2}), respectively. This point is usually overlooked in the literature
\cite{glsu3,mass}. 
  Regarding the sizes of the self-energies we obtain 
\be
\frac{\Sigma_P(\mkq^2;M_Q^2)}{M_P^2}=0.11\pm 0.06~,~0.08\pm 0.04~,0.26\pm 0.06~,
\ee
for pions, kaons and etas, respectively. We then observe that the physical masses of the
 pseudoscalars are dominated by the 
 bare masses and that the dynamical contributions because of their self-interactions  
(which include the 
exchanges of the  scalar resonances $\sigma$, $\kappa$, $f_0(980)$ and $a_0(980)$, as well as of 
 the scalar octet of resonances
around 1.4~GeV) are small. They are however somewhat  more significant 
for the $\eta_8$ ($\mke=545$ MeV from the Gell-Mann-Okubo 
relation). 
We then favour the standard CHPT scenario, where the linear quark 
mass term is assumed to dominate the pseudoscalar masses,  versus the generalized one 
\cite{gchpt}. The $M_{\eta_8}$ that we obtain, around $635$~MeV, is larger than the one of the
 $\eta$ physical meson, $547.45$~MeV, that is very close to the bare $\eta$ mass calculated 
 from Gell-Mann-Okubo. Our value for $M_{\eta_8}$ is  very similar to 
 that of ref.\cite{set}, $M_{\eta_8}=639$~MeV.

Next, we also show our results for the solution of eq.(\ref{eq1}) using the parameters
in the S-waves with their values determined by fitting experimental data employing always 
 physical masses in the interaction kernel. That is,
we do not refit them in terms of the bare masses, as done previously. 
 The resulting numbers  
\be
\mkp=125\pm 4~,~\mkk=477\pm 10~,~M_{\eta_8}= 633 \pm 15 ~\hbox{MeV},
\label{valnox4}
\ee
are  very similar 
to those already determined.
The process of refitting the free parameters in the S-waves by distinguishing between 
physical and bare masses,
 where it corresponds, just introduces minor corrections (compared with the errors quoted)
 and will not be further considered in what follows.

 Taking into account the expression of the bare masses in terms of the quark ones, eq.(\ref{b0mq}), 
and the values  in eq.(\ref{valnox3}),
\be
r_m=\frac{m_s}{\hat{m}}=2\frac{\mkk^2}{\mkp^2}-1=27.1 \pm 2.5~.
\label{valuesrm}\ee
 This number, within errors,
is in the bulk of other determinations,  $r_m=25.7\pm 2.6$ from $\Opc$ CHPT \cite{glsu3},
 or with the more refined one of 
ref.\cite{leutwylerqm}, $r_m=24.4\pm 1.5$, and also with lattice determinations
 \cite{milc}, $27.4 \pm 0.5$. It is worth stressing that our results in this section are predictions without any new free
 parameter. They are a consequence of the strong T-matrices used.

\section{Including  $\Opc$ CHPT self-energies}
\label{sec:opc}

In this 
 section we consider eq.(\ref{eq2}), which  reproduces the 
  CHPT self-energies at $\Opc$ and
  the higher order contributions are estimated  by 
 adding $\Sigma_P^H$. The $\eta-\eta'$ mixing is incorporated 
at ${\cal O}(p^4)$ by the counterterm 
$L_7$ and $M_\eta$ is
fixed to its physical value \cite{glsu3}. Eq.(\ref{eq2}) incorporates the combinations of 
the low energy constants $L_{(5,8)}\equiv 2L_8^r-L_5^r$, $L_{(4,6)}\equiv 2L_6^r-L^r_4$ and 
$L_{(7,8)}\equiv 3L_7+L_8^r$, eqs.(\ref{chptmasses}) 
and (\ref{kchpt}). 
 In table \ref{table:lrzcuadro1} we show  values for 
 these $\Opc$ LECs at the mass of $\rho$ resonance, $\mu=770$~MeV. We show in the second column 
 the so-called main fit  from the ${\cal O}(p^6)$ fits of 
ref.\cite{bijtal}  and in the fifth one those values from the $\Opc$
 fits of ref.\cite{bijopc}. In the fit shown of ref.\cite{bijtal} 
the large $N_c$ suppression of the chiral counterterms $L^r_4$ and $L^r_6$ is used
to fix their values to 0, although the precise scale at which this occurs it is not known. 
 The error bands
in the values of $L^r_4$ and $L^r_6$ in ref.\cite{bijopc}, fifth column,
 are intended to cover this uncertainty. In this
reference the null value for these counterterms is taken at the renormalization 
 scale of the $\eta$ meson mass. 
Other fits in ref.\cite{bijtal} are
also given where this condition is relaxed and the differences are well taken into account by the 
 estimated errors. 
 In the third column we also show the values 
 for $L^r_4$ and $L^r_6$ from
the simultaneous study in ref.\cite{bijkpi} of $\pi\pi$, $K\pi$ scattering 
together with the scalar form
factors, the masses and decay constants of the lightest pseudoscalars,
 and $K_{\ell 4}$ decays at $\Ops$ in CHPT with three flavours. The resulting values 
for such counterterms are found to be small and compatible with zero, in agreement with the large $N_c$
suppression. Note that these new values of 
 ref.\cite{bijkpi}, with  vanishing $L^r_4$ and $L^r_6$ counterterms,  
 reinforce the results of the main fit of ref.\cite{bijtal},  where this was assumed. 
 In refs.\cite{mouss,mouss2} chiral
sum rules are used to fix the latter counterterms to larger values.  These determinations, however, suffer 
 of the generally poor (if any) convergence of the three flavour CHPT series \cite{mass,bijtal,bijkpi},
 with sizable ${\cal O}(p^6)$ contributions. Because  CHPT with
three flavours is used {\it only} at ${\cal O}(p^4)$  the determination of those counterterms 
  incorporates large 
 ${\Ops}$ contributions, which are buried in the numbers given.
   We shall comment more on this issue with a precise example below.
 It should be understood for the third column in table \ref{table:lrzcuadro1}
that when the value for a given $L_i^r$ is not shown in the table this is 
 the same as the one in the second column.

\begin{table}
\begin{center}
\begin{tabular}{lrrrr}
$\times 10^3$	& \cite{bijtal} &  \cite{bijkpi}  &  &  \cite{bijopc}\\
		\hline
$L_4^r$      & $0$       &   $\simeq 0.2$ & $0.2\pm 0.3$ \cite{mouss}    & $-0.3\pm0.5$  	\\
$L_5^r$      & $0.93\pm0.11$  & $-$          &   $-$ & $1.4\pm0.5$	\\
$L_6^r$      & $0$            &  $\lesssim 0$ & $0.4\pm 0.2$ \cite{mouss2}	     & $-0.2\pm0.3$ 	\\
$L_7$      & $-0.31\pm0.14$ &  $-$ & $-$	     &	$-0.4\pm0.2$ 	\\
$L_8^r$      & $0.60\pm0.18$  & $-$ & $-$        &  $0.9\pm 0.3$ \\ \hline
$2L_6^r-L_4^r$ & $0$            & $\simeq -0.2$ & $0.6\pm 0.5$ &  $-0.1\pm0.8$ 	\\ 
$2L_8^r-L_5^r$ & $0.23\pm0.38$ & $-$ &  $-$       & $0.4\pm 0.8$ 	\\
$3L_7+L_8^r$ & $-0.33\pm0.46$  & $-$ &  $-$         &  $-0.3\pm 0.7$  	\\

\hline
\end{tabular}
\caption{Values for some 
 $\Opc$ $\chi$PT counterterms $L_i^r(M_\rho)\times 10^3$ at the renormalization scale 
 of the $\rho$ mass from the literature.  
\label{table:lrzcuadro1}}
\end{center}
\end{table}

We first consider the matching at ${\cal O}(p^4)$ of the self-energies given
 in eqs.(\ref{eq1}) and (\ref{eq2}). The $\pi$ and $K$ self-energies given by eq.(\ref{eq1}) 
 at $\Opc$ are
\ba
M_\pi^2&=&\krig{M}_\pi^2\left\{1+\frac{\krig{M}_\pi^2}{32 \pi^2f^2}\log \frac{\krig{M}_\pi^2}{4\mu^2 e}-
\frac{1}{3}\frac{\krig{M}_\eta^2}{32 \pi^2 f^2}\log \frac{\krig{M}_\eta^2}{4 \mu^2 e}\right\}~,\nn\\
M_K^2&=&\krig{M}_K^2\left\{1+\frac{\krig{M}_\eta^2}{48\pi^2f^2}\log \frac{\krig{M}_\eta^2}{4\mu^2 f^2}\right\}~,
\ea 
with $\log e=1$.  
Comparing this expression with the one  in eq.(\ref{chptmasses}) one easily concludes
that
\be
L_{(4,6)}(\bar{\mu})=-\frac{1}{36}\frac{1}{32\pi^2}\log\frac{\bar{\mu}^2}{4\mu^2e}~~,~~
L_{(5,8)}(\mu)=\frac{1}{6}\frac{1}{32\pi^2}\log\frac{\bar{\mu}^2}{4 \mu^2 e}~,
\label{lsmatched}
\ee
where $\bar{\mu}$ is the renormalization scale in CHPT and $\mu$ was already introduced in eq.(\ref{rmu}).
Notice that the expressions in eq.(\ref{lsmatched}) are ${\cal O}(N_c^0)$, subleading in large $N_c$. 
 This is certainly
expected because this contribution originates from the loops resummed in eq.(\ref{eq1}). Note that $L^r_4$
and $L^r_6$ are ${\cal O}(N_c^0)$, subleading in the large $N_c$ counting,
 while $L^r_5$ and $L^r_8$, separately, are of leading order, ${\cal O}(N_c)$. 
 However, the combination $2L^r_8-L^r_5$ is numerically suppressed, as it is clear from the second
column in table \ref{table:lrzcuadro1} from ref.\cite{bijtal}. In ref.\cite{jaminss} one gets 
the constraint $2L^r_8-L^r_5=0$ by requiring that the strange
scalar form factors vanish in the large $N_c$ limit for $s\to \infty$.

 From eq.(\ref{lsmatched}) it follows that
\be
L_{(5,8)}+6L_{(4,6)}=0~.
\label{conrel}
\ee
This is a robust output  from our main assumption, namely, the dominance of S-wave 
rescattering in the
lightest pseudoscalar self-energies in analogy with the one already well tested for the scalar form factors.
Note that this relation is independent of the renormalization scheme chosen in the calculation 
of our integrals, contrary to the independent values of $2L^r_8-L^r_5$ and $2L^r_6-L^r_4$ in eq.(\ref{lsmatched}).
 E.g. in these equations one has the appearance of the number $e$ in the denominators of the 
 logarithms as a remnant of our choice. This is removed in eq.(\ref{conrel}), which is also
 independent of the renormalization scale $\bar{\mu}$ in CHPT.
 
 One can also extract interesting consequences from eq.(\ref{conrel}). Because of the factor six in
 front of $L_{(4,6)}$ and since $L_{(5,8)}$ is small in modulus,  
 \be
 L_{(4,6)}\simeq 0~.
 \label{l46ours}
 \ee 
 This also implies that 
\be
L^r_6 \simeq \frac{1}{2}L^r_4~, 
\ee
 in good agreement with their most recent estimates in ref.\cite{bijkpi} from $\Ops$ CHPT, third column 
 in table \ref{table:lrzcuadro1}. As commented above, the values  for $L^r_4$ and $L^r_6$ from
 refs.\cite{mouss,mouss2} can suffer from large uncertainties due to the large $\Ops$ 
 contributions, because only CHPT at ${\cal O}(p^4)$ was used. As a
 example, if we take the central values for the bare masses in eq.(\ref{valnox3}), 
 and evaluate with eq.(\ref{chptmasses})  $L_{(4,6)}$ and $L_{(5,8)}$,
\be
L_{(4,6)}=0.23 \cdot 10^{-3}~,~ L_{(5,8)}=0~.
\ee
 These numbers clearly violate the constraint in eq.(\ref{conrel}). 
One would obtain $L_{(5,8)}\simeq -1.3\cdot 10^{-3}$ if the previous value for $L_{(4,6)}$ 
  were used in eq.(\ref{conrel}). This is a clear example that ${\Ops}$ contributions, 
and possibly also higher order ones, are so important 
  in the CHPT series  that unless they are taken proper and explicitly into account, an
  estimation of the $\Opc$ counterterms from phenomenology with only $\Opc$ CHPT is severely biased by
  large higher order contributions buried in the values obtained.

\begin{figure}[ht]
\psfrag{L85}{{\small $2L_8^r-L_5^r$}}
\psfrag{L64}{{\small $2L_6^r-L_4^r$}}
\psfrag{mshat}{{\Large $\frac{m_s}{\hat{m}}$}}
\psfrag{L78}{{\small $3L_7+L^r_8$}}
\centerline{\epsfig{file=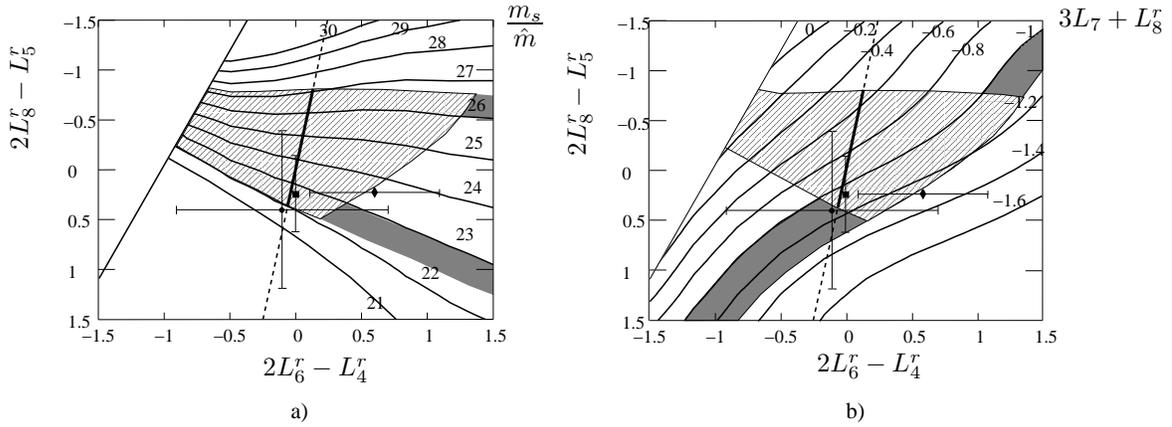,width=.8\textwidth,angle=0}}
\vspace{0.2cm}
\caption{\protect \small Left panel: Contour-plot for $r_m=m_s/\hat{m}$ 
as a function of
$L_{(4,6)}$  and $L_{(5,8)}$. The square corresponds to the second column of table 
\ref{table:lrzcuadro1}, 
the diamond to the fourth one and the circle to the fifth column. The straight line 
corresponds to the relation in eq.(\ref{conrel}). No values for $L^r_5$ nor $L_8^r$ 
are provided in refs.\cite{mouss,mouss2}, and the absence of the vertical 
errorbar in the diamond refers to the fact that $L_{(5,8)}$ cannot be determined then from 
these references. Right panel: 
Contour-plot for $L_{(7,8)}$ as a function of
$L_{(4,6)}$  and $L_{(5,8)}$.  The meaning of the points is the same as in the left  
panel. Our results correspond to the values 
 lying along the straight line within the stripped region, indicated by the thick solid line. 
For more details see the text.} 
\label{fig:lsgrid}
\end{figure}

Next, we apply eq.(\ref{eq2}) to study the physical masses of $\pi$, $K$ and $\eta$. 
  As a function of $L_{(4,6)}$ and $L_{(5,8)}$,
  we solve first for the $\mkp^2$ and $\mkk^2$ bare masses in eq.(\ref{eq2}), and fix 
   $\mke^2$ by the Gell-Mann-Okubo relation, eq.(\ref{okubo}).  $L_{(7,8)}$ is solved from the equation,
\be
M_\eta^2=\mke^2+\Sigma^{4\chi}_\eta(\mkq^2)+\Sigma^H_\eta(\mkq^2;M_Q^2)~,
\label{eqmeta}
\ee
We use the method of eq.(\ref{smi}) for solving eq.(\ref{eq2}), though the last term on the 
r.h.s. of eq.(\ref{smi}) is removed since we are  considering now  
the solution for $\mkp$ and $\mkk$. 
 In fig.\ref{fig:lsgrid}a we show by the contour lines the calculated
values  for $r_m=m_s/\hat{m}$ as a function of
$L_{(4,6)}$  and $L_{(5,8)}$ in units of $10^{-3}$.\footnote{In the whole manuscript, 
the values of the LECs displayed in the figures are given in 
units of $10^{-3}$ and referred to a renormalization scale equal to the $\rho$ mass.} 
The up-left corner where no contour
lines are plotted determines a region where no solutions 
  are found.  The range of values for
$L_{(4,6)}$ and $L_{(5,8)}$ has been chosen to span generously the
values of table~\ref{table:lrzcuadro1} within errors. 
If one takes 
into account that $m_s/\hat{m}=24.4\pm 1.5$  \cite{leutwylerqm},
  the preferred region for  
$L_{(4,6)}$ and $L_{(5,8)}$ would be  between the $23$ and $26$ 
contour lines. The shadowed areas below and above these lines, moving close to the 
22 and 27 contour lines,
represent the uncertainties of
our calculations along them due to the variation in the 
renormalization scale, $\mu\sim
[0.5,1.2]$~GeV, and in the input parameters for the S-waves. 
 In the figure, the square represents the values of $L_{(4,6)}$ and $L_{(5,8)}$ 
  from ref.\cite{bijtal}, second column in table \ref{table:lrzcuadro1}.  
 The full circle corresponds to the values in the last column of the same 
table and the  diamond the value of $L_{(4,6)}$ 
 in the fourth column, refs.\cite{mouss,mouss2}. Notice that no values for $L^r_5$ and $L_8^r$ 
 are  provided in these references and hence no value for $L_{(5,8)}$ can be 
 either determined. This is  
 reflected in the absence of the vertical errorbar 
 for the diamond.
 In the figure it is also shown the line corresponding to the 
 constraint of eq.(\ref{conrel}).  
We show in fig.\ref{fig:lsgrid}b  a contour plot for the
values of $L_{(7,8)}$ that are obtained from eq.(\ref{eqmeta}). 
 If one considers the values  $L_{(7,8)}=-0.33\pm0.46$
\cite{bijtal} at ${\cal O}(p^6)$, and  $L_{(7,8)}=-0.3\pm0.7$ \cite{bijopc}
at ${\cal O}(p^4)$, the preferred $L_{(5,8)}$,
$L_{(4,6)}$ region would be in the interval $\sim [-1,0.4]$. Again the
 shadowed area along these lines
represents the uncertainties of
our calculation.
The same points as in fig.\ref{fig:lsgrid}a are  shown here and 
also the relation of eq.(\ref{conrel}) is depicted by the
straight line. The stripped area in the two figures 
 represents the overlap between the favoured regions on them. Our values lie 
along the thick solid line defined by the intersection of the 
relation in the eq.(\ref{conrel}) with the stripped 
region. It is certainly remarkable the large overlapping between 
this line and the square point from ref.\cite{bijtal}. Note that our approach and that 
of ref.\cite{bijtal} are completely independent. While refs.\cite{bijkpi,bijtal} employ
 resonance saturation hypothesis to estimate the ${\cal O}(p^6)$ counterterms, 
 which has an important impact on their 
NNLO results, we have made used of the dominant role of  
rescattering in meson-meson $I=0$, 1/2 and 1   
scalar dynamics to calculate $\Sigma^H_P$. 
In addition, while refs.\cite{bijkpi,bijtal} consider simultaneously 
  the  lightest pseudoscalar masses and decay constants, 
 $K_{\ell 4}$ decays, scalar form factors and low energy
  $\pi\pi$, $K\pi$ scattering,  we have considered all the scattering channels 
  in the fig.\ref{fig:fitexp} and 
 the $\pi$, $K$ and $\eta$ masses. Taking the intersection of the thick line 
 in fig.\ref{fig:lsgrid}a and the square  from ref.\cite{bijtal} at the 
 one $\sigma$ level, we end with the preferred interval of values 
 \be
 -0.15\lesssim L_{(5,8)}\cdot 10^3\lesssim   0.35~,
 \label{intval}
  \ee
 for $\bar{\mu}=M_\rho$. Let us recall also 
 our previous estimate for $L_{(4,6)}\simeq 0$ in eq.(\ref{l46ours}).

We end this section with a remark. 
 One can also use eq.(\ref{lsmatched}) to derive values independently for
$L_{(5,8)}$ and $L_{(4,6)}$. Nevertheless, these values are affected by the choice 
of the renormalization scheme, so that they can differ from those in the standard modified 
$\overline{MS}$ scheme of CHPT. In addition, allowing $\mu$ to change 
within the range $0.5-1.2$~GeV the range of values of $L_{(5,8)}$ is rather large and 
does not add any real new restriction to what it is already shown in fig.\ref{fig:lsgrid}.
Notice that when we consider eq.(\ref{eq2}) the ${\cal O}(p^4)$ contribution is given by
$\Sigma_P^ {4\chi}$, so that the $L^r_i$ are considered properly in the same renormalization
scheme as in CHPT. The same can be said for the relation in eq.(\ref{conrel}) as it is
 invariant under any new constants that could be added when removing the infinite part to
 calculate the integral in eq.(\ref{onehas}).

\section{Comparison with lattice QCD results}
\label{lattice}

\begin{figure}[ht]
\psfrag{L85}{$2L_8^r-L_5^r$}
\psfrag{L64}{$2L_6^r-L_4^r$}
\psfrag{mshat}{{\Large $\frac{m_s}{\hat{m}}$}}
\centerline{\epsfig{file=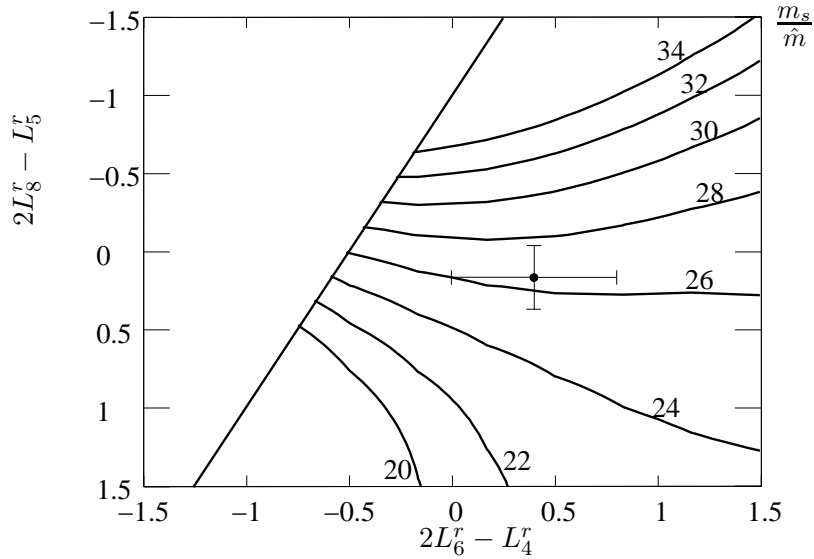,width=.6\textwidth,angle=0}}
\vspace{0.2cm}
\caption{\protect {\small Contour-plot for $r_m=m_s/\hat{m}$ 
as a function of
$L_{(4,6)}$  and $L_{(5,8)}$ obtained by employing only the $\Opc$ CHPT self-energies. 
 The point 
corresponds to the values of $L_{(5,8)}$ and $L_{(4,6)}$ from the MILC collaboration \cite{milc1}.}}
\label{fig:rm_onlyp4}
\end{figure}  
It is worth comparing the result in fig.\ref{fig:lsgrid}a with the
calculation obtained by considering  the self-energies only up to
${\cal O}(p^4)$ in CHPT, shown in 
 fig.\ref{fig:rm_onlyp4}. The difference in $m_s/\hat{m}$ for the same $L_{(4,6)}$ and 
 $L_{(5,8)}$ is typically around 
 $1-3$ units between both calculations, with larger values when only the ${\cal O}(p^4)$ CHPT self-energies 
are taken into account. A similar trend was also observed in ref.\cite{mass} from pure CHPT to $\Ops$,
 with  smaller values as
 higher orders in the chiral series are included. To $\Opc$ the values 24.9 and 24.4 
 for $r_m$ become, respectively,  24.1 and 23.3 to ${\cal O}(p^6)$ \cite{mass}.

 Two loop diagrams  contributing to the self-energies 
 were  found very relevant numerically in ref.\cite{mass}, larger by several factors than the ${\cal
 O}(p^4)$ contributions, similarly to what we find. Since diagrams with more than one loop 
  were neglected in refs.\cite{milc1,milc},  worries arise about 
 systematic uncertainties hidden in the  values determined in these references  for low 
  energy parameters. To illustrate this point, 
  we show in fig.\ref{fig:rm_onlyp4} by the solid circle the values $L_{(5,8)}=(0.16\pm 0.20)\cdot 10^{-3}$ and 
 $L_{(4,6)}=(0.4\pm 0.4)\cdot 10^{-3}$ given by the QCD lattice calculation of the MILC Collaboration \cite{milc1}. 
 The latter reference also predicts 
 $m_s/\hat{m}=27.4\pm 0.4$. It is worth stressing  that this  value is very close to the one
 obtained by considering only  $\Opc$ CHPT, as shown in the figure. 
 Note that ${\cal O}(p^4)$ CHPT  also includes at most diagrams with one 
  chiral loop. As a result, we find quite reasonable to expect 
 a reduction in the value of $m_s/\hat{m}$ of one unit at least, 
 if higher loop diagrams were included, as we have done. Certainly, further arguments to those given 
 in ref.\cite{milc1} are necessary to rule out this expectation. 

The MILC Collaboration also provides results for the pseudoscalar meson masses. 
We present next a good fit to these lattice data on the pseudoscalar masses ($\pi$ and $K$), 
at the level of 1$\xxpc$,
 making use of  eq.(\ref{eq2}) for the self-energies. 
  We are aware that our parameterization 
   does not take into account the taste symmetry violating effects, 
   due to the finite lattice spacing. Nevertheless, we think that our fit is valuable since 
   the differences     between our full results
  and those  from a fit to lattice data with only ${\cal O}(p^4)$ CHPT, eq.(\ref{chptmasses}),  
 is an estimate of these diagrams with a higher number of loops. 
 It provides then a  rough quantification 
 of systematic uncertainties that could affect the values 
 for $L_{(4,6)}$, $L_{(5,8)}$ and $m_s/\hat{m}$ from refs.\cite{milc,milc1,ultimo}.
  It is true that
 these references include ${\cal O}(p^6)$ contributions at tree level through the
  appropriate chiral
 counterterms in Staggered CHPT. Nevertheless, since  two-loop diagrams, as well 
 as one loop diagrams
 with higher order vertices, are neglected, the procedure is not systematic and 
 some quantification is of interest.
   
   We employ our eq.(\ref{eq2}) to
reproduce the quark mass dependence of the pseudoscalar masses provided by the MILC Collaboration
\cite{milc1}  with three dynamical quarks  
 and equal valence and sea quark masses.  
 The systematic procedure to extrapolate these lattice results
  to the lightest quark masses in the continuum 
 is to employing Staggered CHPT \cite{9l,apco} with three flavours. Nevertheless,
  present applications suffer  of a huge proliferation of free parameters and  
 include only  diagrams involving at most one loop, though
   supplied with tree diagrams up to NNLO \cite{milc1}. 

\begin{figure}[ht]
\psfrag{K}{$M_K$}
\psfrag{P}{$M_\pi$}
\psfrag{mshat}{$m_s/\hat{m}$}
\psfrag{phys}{{\tiny physical point}}
\psfrag{pi}{$\pi$}
\centerline{\epsfig{file=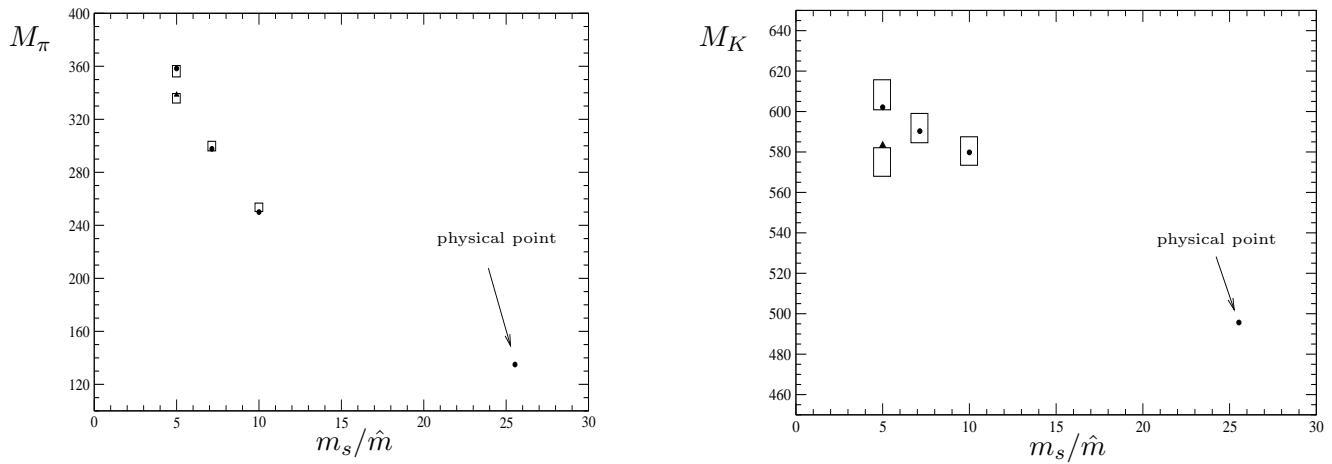,height=6cm,width=1.\textwidth,angle=0}}
\vspace{0.2cm}
\caption{\protect {\small Lattice data from the MILC Collaboration \cite{milc1} 
and their reproduction by our parameterization. The masses are given in MeV. 
The size of the squares corresponds to a 
relative error of $1.2$\%, the one given to $a$ in ref.\cite{milc1}. The circles refer to our 
calculations for the coarse lattices and the triangles for the fine ones. The physical mass values 
are signalled out by an arrow.}}
\label{fig:masseslat}
\end{figure}

\begin{figure}[ht]
\psfrag{L85}{$2L_8^r-L_5^r$}
\psfrag{L64}{$2L_6^r-L_4^r$}
\psfrag{mshat}{{\Large $\frac{m_s}{\hat{m}}$}}
\centerline{\epsfig{file=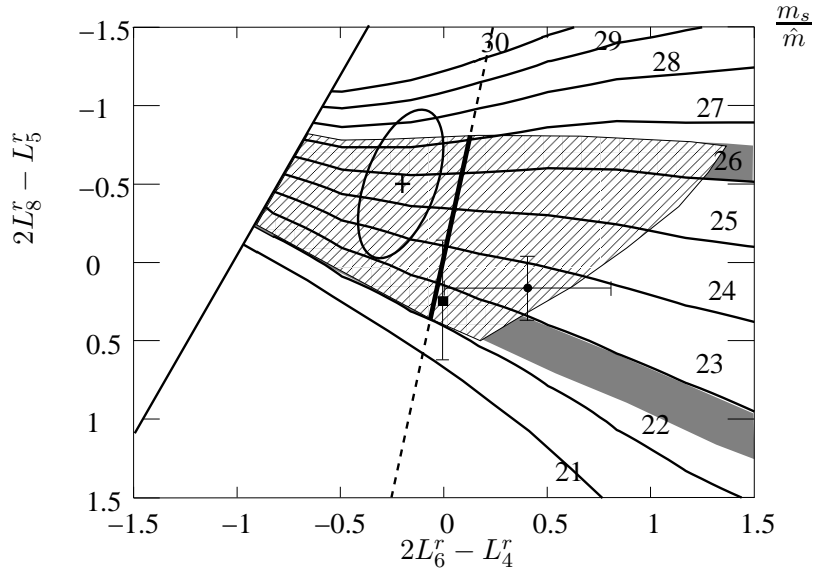,width=.6\textwidth,angle=0}}
\vspace{0.2cm}
\caption{\protect {\small Contour-plot for $r_m=m_s/\hat{m}$ 
as a function of
$L_{(4,6)}$  and $L_{(5,8)}$}, showing also the ellipse  of 
the fitted values in eq.(\ref{l58l46lat}). The circle corresponds 
to the lattice extrapolation from Staggered CHPT \cite{milc1}. The rest is the same 
as in fig.\ref{fig:lsgrid}a.}
\label{fig:rmlattice}
\end{figure}

    The strategy is the following. The MILC Collaboration \cite{milc1} provides
pseudoscalar meson masses as a function of the bare quark masses used. 
 The current quark masses and the bare ones   are 
proportional \cite{milc}, 
\be
\krig{M}_\pi^2=2B_0 C(a) a\hat{m}~,~~\krig{M}_K^2=B_0 C(a) a(\hat{m}+m_s)~,
~~\krig{M}_\eta^2=\frac{2}{3}B_0 C(a) a(\hat{m}+2m_s)~.
\ee
The dependence of the constant $C(a)$ with the lattice spacing 
 can be found in ref.\cite{ultimo} at the 
two loop level, and in ref.\cite{milc} calculated up to the one loop order. 
 Previous calculations with $C(a)$ at the one loop level \cite{milc} gave a value 
for $m^{\overline{MS}}_s(2 \hbox{GeV})=76\pm 8$ MeV, to be compared with the latest 
value given in ref.\cite{ultimo} $m^{\overline{MS}}_s(2 \hbox{GeV})=87\pm 6$ MeV. This substantial 
shift upwards comes from the improvement in the determination of $C(a)$ by moving
 from the one to the two loop calculation.  
We use in our fits the   two loop result for $C(a)$ from ref.\cite{ultimo}.
 Ref.\cite{milc1} gives the so called ``coarse" lattice runs with $a_{coarse}\simeq 0.12$~fm, 
 and the ``fine" lattices with $a_{fine}\simeq 0.09$~fm, with an error in $a$ about $1.2\xxpc$. 
Making use of ref.\cite{ultimo} we then determine that $
C(a_{fine})/C(a_{coarse})=2.76/1.85=1.49~$ 
for $\mu=2$~GeV. From this relation we fix $C(a_{fine})$ in terms of $C(a_{coarse})$ and 
take the latter  as a free parameter, denoted  in the following just as $C$. 
  As the constant $C$ does not depend 
on the value of the quark mass in very good numerical accuracy (at the level
of $0.1$\% \cite{milc,ultimo}), we then employ the same value for all the quark masses
 used in the  lattice runs with the same $a$. 
We  solve
 eq.(\ref{eq2}) for the kaon and pion masses given by lattice QCD
  in terms of the bare quark masses used. From the mass equations 
 at the physical point one can obtain
 $L_{(4,6)}$ and $L_{(5,8)}$ as a function of 
 $\mkp^2$ and $\mkk^2$. Hence, the only free parameters in the fit
to lattice data are  $B_0 C$, $\mkp^2$ and $\mkk^2$. In the fit we do not consider 
the two points of the fine lattice run with $m_s/\hat{m}=0.031/0.0124=2.5$, because it 
implies a rather heavy pion with a mass of 470 MeV \cite{milc1}.
We then restrict ourselves to the case of lighter pions with $m_s/\hat{m}\geq 5$. 
In figs.\ref{fig:masseslat}a, b, we show the reproduction of the lattice
points \cite{milc1}  
 and of the physical values for the pion and kaon masses, respectively.
 The size of the squares corresponds to a relative error of 1.2\%, the one given to $a$ \cite{milc1}. 
The circles correspond to our points for the coarse lattices, the triangles for the 
fine ones and the arrows indicate the values of the physical kaon and pion masses. 
 It is also worth stressing  that   the scattering data  in fig.\ref{fig:fitexp} 
 are simultaneously reproduced.\footnote{ In the fit to
the unphysical points of lattice the dependence on the bare masses
of the pseudoscalar decay
constants $f_\pi$, $f_K$  and $f_\eta$, generally called $f_P$, could play a role. We have
then also performed fits to the lattice data recalculating the $f_P$ in terms of the $\Opc$
CHPT expressions for $f_P$ as a function of the bare masses, with values 
for $L^r_4$ and $L^r_5$ such that  at the physical point the constants $f_P$  have 
 their physical values. As we have checked, these considerations affect little  the
 resulting fit and the results are  well within the uncertainty
bounds quoted before. } From this fit we get the
values,
\be
2L_8^r-L_5^r=-0.52\pm0.43\quad,\quad
~2L^r_6-L_4^r=-0.20\pm0.17~.
\label{l58l46lat}
\ee
 In fig.\ref{fig:rmlattice} we show by the solid ellipse the one sigma region 
for $L_{(4,6)}$ and $L_{(5,8)}$ obtained in our fit, 
from where one can infer the correlation between these two parameters. 
We also show the straight line from eq.(\ref{conrel}), the square point corresponding 
to ref.\cite{bijtal}, second column in table \ref{table:lrzcuadro1}, and 
by the circle we show the values 
\be
2L_8^r-L_5^r=0.16\pm0.20\quad,\quad
~2L^r_6-L_4^r=0.4\pm0.4~,
\label{milcmilc}
\ee
obtained by the MILC Collaboration \cite{milc,milc1}. 
The fact that the values in eq.(\ref{l58l46lat}) do not lie within the favoured region in 
 fig.\ref{fig:rmlattice} is an indication of non-negligible finite lattice spacing effects,  
 already stressed in ref.\cite{milc1}. The MILC values in eq.(\ref{milcmilc}) are 
 barely compatible, at the level of one sigma, with the thick solid line shown inside 
 the stripped region and with the square from   ref.\cite{bijtal}.

   We have also performed a fit employing only the
self-energies calculated at $\Opc$ in CHPT. The result obtained from
 the fit to the lattice data is  
$2L_8^r-L_5^r=-0.20\cdot 10^{-3}$ and  
$2L^r_6-L_4^r=0.13\cdot 10^{-3}$ and these data is quite well reproduced. 
The difference with respect to eq.(\ref{l58l46lat}) 
is due to the extra unitarity loops and 
higher order effects from the exchanges of tree level resonances.  The former effects 
are not systematically considered in ref.\cite{milc1}. The size of 
this difference is then an estimate 
of systematic uncertainties that could affect those results from ref.\cite{milc1} and that, when 
added in quadrature with the errors given, 
would enlarge by a factor $\sim 2$ the errorbar for $L_{(5,8)}$ in eq.(\ref{milcmilc}).

 From the favoured region in fig.\ref{fig:lsgrid}a, given by the intersection of 
 the straight line with the stripped region and the square point within errorbars, it results
\be
\frac{m_s}{\hat{m}}=22-25~.
\label{rmlat}
\ee
This value is in good agreement with $24.4\pm 1.5$ from ref.\cite{leutwylerqm}. 
The previous result is somewhat lower 
 than  the value $27.4\pm 0.5$ from the MILC Collaboration \cite{milc} and ref.\cite{ultimo}.
 As discussed above, we consider that this difference is mostly due to the non inclusion of higher loop 
 diagrams in the Staggered CHPT extrapolations used in these references. It follows from this 
 discussion that those calculations are necessary to settle this issue. 
  Within the same favoured  region of values for $L_{(4,6)}$ and $L_{(6,8)}$ we also obtain, 
\be
 L_{(7,8)}=-(0.7-1.1)\cdot 10^{-3}~,
\ee
and
\be
\label{bmlat}
\mkp=116-119~,~ \mkk= 400-420 ~\hbox{ MeV}.
\ee
Finally, we give in table~\ref{tab:ratioS} the relative sizes of $\Sigma^{4\chi}_P$ and 
$\Sigma^H_P$ for the same interval of values for $L_{(5,8)}$, eq.(\ref{intval}),
 and $L_{(4,6)}=0$.
 The fact that  $\Sigma^H_P$ is larger than $\Sigma^{4\chi}_P$
is in agreement with ref.\cite{bijtal}. Indeed, 
the sizes of our self-energies from $\Sigma^H_P$ are rather similar
to those determined in this reference to ${\cal O}(p^6)$.
Ref.\cite{mass} obtains:
$\Sigma^{6\chi}_\pi/M_\pi^2=0.132-0.355$, 
$\Sigma^{6\chi}_K/M_K^2=0.194-0.423$ and
$\Sigma^{6\chi}_\eta/M_\eta^2=0.234-0.521$. These values are  well
inside our bulk of results in table \ref{tab:ratioS}.  Thus, the
calculations of ref.\cite{bijtal} to ${\cal O}(p^6)$, although
showing that this order is much larger than the ${\cal O}(p^4)$,
does not imply necessarily  the lack of convergence of the chiral
series. Our calculation, estimating higher orders corrections 
by incorporating physical S-waves which include both resonant and
non-resonant dynamics, gives  us values of similar size to those
of this reference up to ${\cal O}(p^6)$. In addition, we have to recall the 
agreement between the straight line in fig.\ref{fig:lsgrid}a, determined from an algebraic 
matching at ${\cal O}(p^4)$ of eqs.(\ref{eq1}) and (\ref{eq2}),  and the square 
from ref.\cite{bijtal}, determined numerically from a phenomenological analysis to 
$\Ops$.

\begin{table}
\begin{center}
\begin{tabular}{|c|c|c|c|}
\hline
 	&     $\pi$       &       $K$      &           $\eta $\\
\hline
$\Sigma^{4\chi}_P/M_P^2 $ & $[-0.010,0.015]$&$[-0.01,-0.07]$ & $[-0.16,-0.18]$ \\
$\Sigma^H_P/M_P^2$ &$[0.233,0.235]$ &$[0.352,0.358]$ &$[0.445,0.460]$ \\
$\Sigma^{4\chi}_P/\Sigma^H_P $ & $[-0.05,0.06]$ & $[-0.03,-0.20]$ & $[-0.37,-0.40]$\\
\hline
\end{tabular}
\caption{Relative sizes of the $\Opc$,  ${\cal O}(p^6)$ and higher 
order contributions to the self-energies. 
The intervals corresponds to the interval of values for $L_{(5,8)}$ in eq.(\ref{intval}) 
and $L_{(4,6)}=0$. 
 \label{tab:ratioS}}
\end{center}
\end{table}

\section{Conclusions}
\label{sec:con}

We have undertaken a non-perturbative chiral  study of the self-energies of the
lightest pseudoscalar mesons.  In their evaluation the S-wave amplitudes obtained in UCHPT, that 
reproduce scattering data very accurately up to around 1.4~GeV, are employed. These amplitudes generate 
the exchange of the nonet of the lightest scalar resonances, 
$\sigma$, $f_0(980)$, $a_0(980)$ and $\kappa$,
whose contributions to the pseudoscalar masses are expected to be relevant since they
 significantly affect the scalar form factors \cite{gasmeis,scalarpi,orocaff,yff}. The latter are
 tightly related to the pseudoscalar self-energies due to the Feynman-Hellman theorem. As stressed, their
 contributions have to be calculated simultaneously with 
 those from the unitarity loops, since these resonances due their 
 origin to the large unitarity contributions driven by chiral symmetry.

We determine that the pseudoscalar masses squared
 for $\pi$  and $K$  are dominated by the bare ones, with the self-energy contributions not larger 
 than 10$\xxpc$. For the $\eta$ meson they can be larger but still lower than 
 around $30\xxpc$. This favors standard CHPT versus its generalized scenario. We also determine 
 the mass of the $\eta_8$, $M_{\eta_8}=635$ MeV, in agreement with a previous determination. Next, 
 we take the exact $\Opc$ CHPT self-energies and resum the higher order contributions 
within our approach. Matching algebraically 
at $\Opc$ with the  calculated self-energies from CHPT one obtains 
the relation $L_{(5,8)}=-6L_{(4,6)}$, which is independent of the renormalization scale and 
of our choice of renormalization scheme. From this relation it follows that $L_{(4,6)}\simeq 0$ and 
thus $L^r_6\simeq L^r_4/2$, 
in good agreement with the $\Ops$ CHPT results from ref.\cite{bijtal}. 
Considering as well the value of the quark mass ratio 
$m_s/\hat{m}=24.4\pm 1.5$ from ref.\cite{leutwylerqm}, and 
 the values for $L_{(7,8)}$ from refs.\cite{bijtal,bijopc}, we delimit a shorter interval 
 of values for $L_{(5,8)}$ lying along the line $L_{(5,8)}+6L_{(4,6)}=0$. 
The value obtained from refs.\cite{bijkpi,bijtal} perfectly overlaps with our 
determination, and the common intersection between these two independents methods 
gives $-0.15\lesssim L_{(5,8)}\cdot 10^3\lesssim 0.35$. 
   This implies that we estimate $m_s/\hat{m}=22-25$ and $L_{(7,8)}=-(0.7-1.1)\cdot 10^{-3}$.
    We also warn 
about higher loop diagram effects in the chiral series when evaluating  $m_s/\hat{m}$, as they 
tend to decrease the value for this ratio. This is relevant for its recent calculation 
 from lattice QCD \cite{milc1} as these higher loop effects are not 
 systematically accounted for in present extrapolations to physical light quark masses 
 and the continuum. Their evaluation is called for. 
  We have also offered a good reproduction of the lattice pseudoscalar
 masses, though the fitted values 
 for $L_{(4,6)}$ and $L_{(5,8)}$ are outside the previous favoured region. 
 This is considered as a clear indication of discretization effects not yet negligible in the 
 lattice data, as already stressed in ref.\cite{milc1}. 
 
Other interesting result from our investigation is that although the $\Ops$ contribution to 
the self-energies calculated in three flavour CHPT \cite{mass} is much larger than the  $\Opc$ ones,
 we still obtain 
self-energies of similar values to those of ref.\cite{mass} despite higher 
order corrections are also included.
 This fact together with  our agreement with ref.\cite{bijtal} in the values for 
$L_{(4,6)}$ and $L_{(5,8)}$, indicate that the SU(3) CHPT expansion
  is not spoiled  once 
the $\Ops$ is taken into account.

\section*{Acknowledgements}
 Financial support by MEC (Spain)  grants 
 No. FPA2004-62777, No. FPA2007-62777, 
 No. BFM-2003-00856, Fundaci\'on S\'eneca (Murcia) grant Ref. 02975/PI/05,
  the European Commission (EC)  RTN Program Network ``EURIDICE'' 
 Contract No. HPRN-CT-2002-00311 and the HadronPhysics I3
Project (EC)  Contract No. RII3-CT-2004-506078 is acknowledged.


\end{document}